\documentclass[sigconf]{acmart}

\usepackage{booktabs}
\usepackage{multirow}
\usepackage{footnote}
\usepackage{bm}
\usepackage{subcaption}
\usepackage{siunitx}
\usepackage{mathtools}  
\usepackage{dblfloatfix}    

\usepackage{scalerel}
\usepackage{tikz}
\usetikzlibrary{svg.path}

\definecolor{orcidlogocol}{HTML}{A6CE39}
\tikzset{
  orcidlogo/.pic={
    \fill[orcidlogocol] svg{M256,128c0,70.7-57.3,128-128,128C57.3,256,0,198.7,0,128C0,57.3,57.3,0,128,0C198.7,0,256,57.3,256,128z};
    \fill[white] svg{M86.3,186.2H70.9V79.1h15.4v48.4V186.2z}
                 svg{M108.9,79.1h41.6c39.6,0,57,28.3,57,53.6c0,27.5-21.5,53.6-56.8,53.6h-41.8V79.1z M124.3,172.4h24.5c34.9,0,42.9-26.5,42.9-39.7c0-21.5-13.7-39.7-43.7-39.7h-23.7V172.4z}
                 svg{M88.7,56.8c0,5.5-4.5,10.1-10.1,10.1c-5.6,0-10.1-4.6-10.1-10.1c0-5.6,4.5-10.1,10.1-10.1C84.2,46.7,88.7,51.3,88.7,56.8z};
  }
}

\newcommand\orcidicon[1]{\href{https://orcid.org/#1}{\mbox{\scalerel*{
  \begin{tikzpicture}[yscale=-1,transform shape]
  \pic{orcidlogo};
  \end{tikzpicture}
}{|}}}}

\DeclarePairedDelimiter\abs{\lvert}{\rvert}%
\DeclarePairedDelimiter\norm{\lVert}{\rVert}%
\makeatletter
\let\oldabs\abs
\def\abs{\@ifstar{\oldabs}{\oldabs*}}
\let\oldnorm\norm
\def\norm{\@ifstar{\oldnorm}{\oldnorm*}}
\makeatother

\newcommand{\ra}[1]{\renewcommand{\arraystretch}{#1}}

\copyrightyear{2019}
\acmYear{2019}
\setcopyright{acmlicensed}
\acmConference[NICE '19]{Neuro-inspired Computational Elements Workshop}{March 26--28, 2019}{Albany, NY, USA}
\acmBooktitle{Neuro-inspired Computational Elements Workshop (NICE '19), March 26--28, 2019, Albany, NY, USA}
\acmPrice{15.00}
\acmDOI{10.1145/3320288.3320303}
\acmISBN{978-1-4503-6123-1/19/03}

\hypersetup{final}
\begin{document}

\title{Analysis of Wide and Deep Echo State Networks for Multiscale Spatiotemporal Time Series Forecasting}



\author{Zachariah Carmichael, Humza Syed, Dhireesha Kudithipudi}
\orcid{0000-0002-7603-2004}
\orcid{0000-0003-4462-5224}
\affiliation{%
  \institution{Neuromorphic AI Lab, Rochester Institute of Technology}
  \streetaddress{1 Lomb Memorial Dr.}
  \city{Rochester}
  \state{New York}
  \postcode{14623}
}
\email{{zjc2920, hxs7174, dxkeec}@rit.edu}

\renewcommand{\shortauthors}{Z. Carmichael \textit{et al.}}
\renewcommand{\shorttitle}{Analysis of Wide and Deep ESNs for Multiscale Spatiotemporal Time Series Forecasting}

\begin{abstract}

Echo state networks are computationally lightweight reservoir models inspired by the random projections observed in cortical circuitry. As interest in reservoir computing has grown, networks have become deeper and more intricate. While these networks are increasingly applied to nontrivial forecasting tasks, there is a need for comprehensive performance analysis of deep reservoirs. In this work, we study the influence of partitioning neurons given a budget and the effect of parallel reservoir pathways across different datasets exhibiting multi-scale and nonlinear dynamics.

\end{abstract}

%
%


\maketitle

\section{Introduction}

Recurrent neural networks (RNNs) have recently shown advances on a variety of spatiotemporal tasks due to their innate feedback connections. One of the most commonly used RNNs today is the long short term memory (LSTM) network \cite{hochreiter1997long}. LSTMs are widely employed for solving spatiotemporal tasks and demonstrate high accuracy. However, these networks are generally prone to expensive computations that often lead to long training times.

A computationally lightweight approach to address spatiotemporal processing is reservoir computing (RC), which is a biologically-inspired framework for neural networks. RC networks comprise hidden layers, referred to as reservoirs, that consist of pools of neurons with fixed random weights and sparse random connectivity. Echo state networks (ESNs) \cite{jaeger2001echo} and liquid state machines (LSMs) \cite{maass_LSM_2004} are the two major types of RC. Both architectures make use of sparse random connectivity between neurons to mimic an intrinsic form of memory, as well as enable rapid training, as training occurs only within the readout layer. Whereas ESNs are rate-based models, liquid state machines (LSMs) are spike-based. The focus of this work is primarily on ESNs.

\begin{figure}
    \centering
    \includegraphics[width=\linewidth]{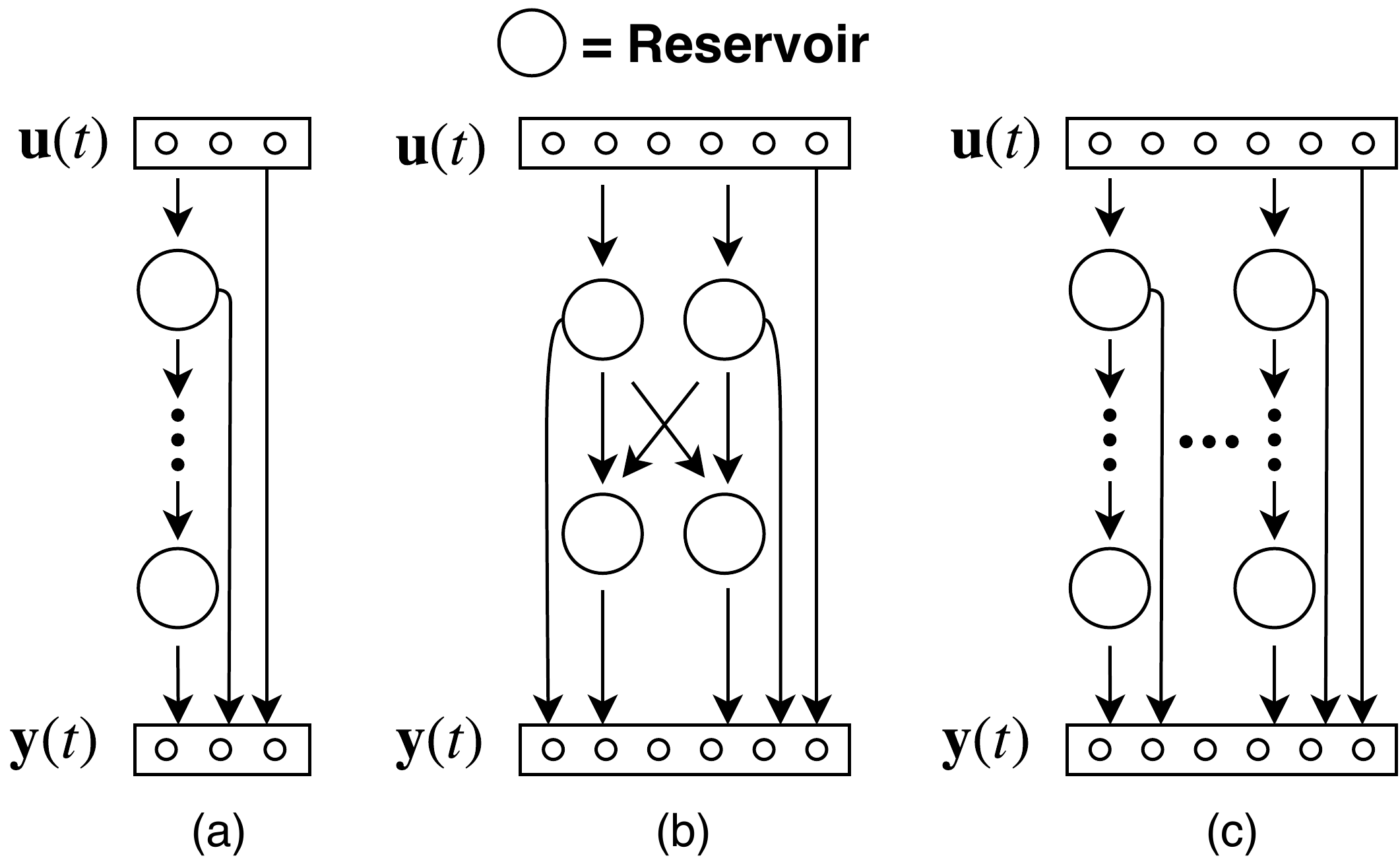}
    \caption{Various \emph{Mod-DeepESN} topologies.}
    \label{fig:topologies}
\end{figure}

In general, ESNs have been shown to perform well on small spatiotemporal tasks but underperform as task complexity increases. Prior literature has shown that ESNs are capable of various functions, such as speech processing, EEG classification, and anomaly detection \cite{soures2017, jaeger_optimization_2007}. In recent literature, several groups have begun to study how these networks can cope with increasingly complex time series tasks with dynamics across multiple scales and domains \cite{maDeepESNMultipleProjectionencoding2017, mesm_2017, r2sp_2013, gallicchio2018design}. One technique to enhance ESNs is the addition of reservoir layers. These networks are referred to as deep ESNs, which provide a hierarchical framework for feature extraction and for capturing nontrivial dynamics while maintaining the lightweight training of a conventional ESN. \citeauthor{maDeepESNMultipleProjectionencoding2017} introduced the Deep-ESN architecture which utilizes a sequence of multiple reservoir layers and unsupervised encoders to extract intricacies of temporal data \cite{maDeepESNMultipleProjectionencoding2017}. \citeauthor{gallicchio_deep_2017} proposed an architecture, named DeepESN, that utilizes \citeauthor{jaeger_optimization_2007}'s leaky-integrate neurons in a deeper ESN architecture \cite{gallicchio_echo_2017, jaeger_optimization_2007}.

In our previous work, we introduced the \emph{Mod-DeepESN}, a modular architecture that allows for varying topologies of deep ESNs \cite{Mod-DeepESN}. Intrinsic plasticity (IP) primes neurons to contribute more equally towards predictions and improves the network's performance. A network with a wide and layered topology was able to achieve a lower root-mean-squared error (RMSE) than other ESN models on a daily minimum temperature series \cite{temp_data} and the Mackey-Glass time series. In this paper, we propose alternative design techniques to enhance the \emph{Mod-DeepESN} architecture. Comprehensive analysis is required to understand why these networks perform well.

\section{Architecture}

A framework under the RC paradigm, ESNs are RNNs which comprise one or more reservoirs with a population of rate-based neurons. Reservoirs maintain a set of random untrained weights and exhibit a nonlinear response when the network is driven by an input signal. The state of each reservoir is recorded over the duration of an input sequence and a set of output weights is trained on the states based on a teacher signal. As the output computes a simple linear transformation, there is no need for expensive backpropagation of error throughout a network as is required for training RNNs in the deep learning paradigm. Thus, the vanishing gradient problem is avoided while still being able to capture complex dynamics from the input data. Vanilla ESNs comprise a single reservoir and have limited application, especially with data exhibiting multi-scale and highly nonlinear dynamics. To this end, various architectures have been proposed with multiple reservoirs, additional projections, autoencoders, plasticity mechanisms, etc. \cite{r2sp_2013, maDeepESNMultipleProjectionencoding2017, mesm_2017, gallicchio_deep_2017, gallicchio2018design, Mod-DeepESN}.

Building off of \cite{gallicchio_deep_2017, Mod-DeepESN}, we introduce the flexible \emph{Mod-DeepESN} architecture, which maintains parameterized connectivity between reservoirs and the input. A broad set of topologies are accommodated by its modularity, several of which are shown in Figure \ref{fig:topologies}. We denote the tensor of input data $\mathcal{U} \in \mathbb{R}^{N_S \times N_t \times N_U}$ which comprises $N_S$ sequences of $N_t$ timesteps and $N_U$ features. $N_t$ may differ between each of the $N_S$ sequences, but for simplicity such variability is left out of the formulation. Reservoirs that are connected to the input receive the vector $\mathbf{u}(t) \in \mathbb{R}^{N_U}$ at timestep $t$ where $\mathbf{u}(t) \in \mathbf{U} \in \mathbb{R}^{N_t \times N_U}$ and $\mathbf{U} \in \mathcal{U}$.
$\mathbf{u}(t)$ is mapped by the input weight matrix $\mathbf{W}_{in} \in \mathbb{R}^{N_U \times \left\lVert \mathbf{C}_\mathbf{u} \right\rVert_0 N_R}$ into each reservoir. $N_R$ is the number of neurons per reservoir and typically $N_R \gg N_U$. The binary matrix $\mathbf{C}$ determines the feedforward connections between reservoirs and the input $\mathbf{u}$. For example, if element $\mathbf{C}_{\mathbf{u}, 2}$ is `1', then $\mathbf{u}$ and reservoir 2 are connected. $\left\lVert \mathbf{C}_\mathbf{u} \right\rVert_0$ gives the number of reservoirs that are connected to $\mathbf{u}$. The output of the $l^{\textup{th}}$ reservoir, $\mathbf{x}^{(l)}\in\mathbb{R}^{N_R}$, is computed as \eqref{eq:layer_output}
\begin{subequations}\label{eq:layer_output}
\begin{align}
    \begin{split}\label{eq:layer_output_a}
        \tilde{\mathbf{x}}^{(l)}(t)=&~ \tanh\left( \mathbf{W}_{res}^{(l)}\mathbf{i}^{(l)}(t)+ \hat{\mathbf{W}}_{res}^{(l)}\mathbf{x}^{(l)}(t-1) \right)
    \end{split}\\
    \begin{split}\label{eq:layer_output_b}
        \mathbf{x}^{(l)}(t)=&~ (1-a^{(l)})\mathbf{x}^{(l)}(t-1)~+a^{(l)}\tilde{\mathbf{x}}^{(l)}(t)
    \end{split}
\end{align}
\end{subequations}

\noindent where $\mathbf{i}^{(l)}(t)$ is given by \eqref{eq:layer_input}.
\begin{equation}\label{eq:layer_input}
    \mathbf{i}^{(l)}(t) =
    \begin{cases}
        \mathbf{u}(t) & l=1 \\
        \mathbf{x}^{(l-1)}(t) & l>1 \\
    \end{cases}
\end{equation}

\noindent$\mathbf{W}_{res}^{(l)}\in\mathbb{R}^{N_R\times N_R}$ is a feedforward weight matrix that connects two reservoirs, while $\hat{\mathbf{W}}_{res}^{(l)}\in\mathbb{R}^{N_R\times N_R}$ is a recurrent weight matrix that connects intra-reservoir neurons. The per-layer leaky parameter $a^{(l)}$ controls the leakage rate in a moving exponential average manner. Note that the bias vectors are left out of the formulation for simplicity. The state of a \emph{Mod-DeepESN} network is defined as the concatenation of the output of the $N_L$ reservoirs, i.e. $\mathbf{x}(t)=(\mathbf{u}(t), \mathbf{x}^{(1)}(t),...,\mathbf{x}^{(N_L)}(t))\in\mathbb{R}^{N_U + N_L N_R}$. The matrix of all states is denoted as $\mathbf{X}=(\mathbf{x}(1),\mathbf{x}(2),...,\mathbf{x}({N_t}))\in\mathbb{R}^{N_S N_t \times (N_U + N_L N_R)}$. Finally, the output of the network for the duration $N_t$ is computed as a linear combination of the state matrix using \eqref{eq:output}.
\begin{equation}\label{eq:output}
    \mathbf{Y}=\mathbf{X}\mathbf{W}_{out}
\end{equation}

\noindent The matrix $\mathbf{W}_{out} \in \mathbb{R}^{(N_U + N_L N_R) \times N_Y}$ contains the feedforward weights between reservoir neurons and the $N_Y$ output neurons, and $\mathbf{Y} \in \mathbb{R}^{N_S N_t \times N_Y}$ is the ground truth with a label for each timestep. In a forecasting task, the dimensionality of the output is the same as the input, i.e. $N_Y = N_U$.
Ridge regression, also known as Tikhonov regularization, is used to solve for optimal $\mathbf{W}_{out}$ and is shown with the explicit solution in \eqref{eq:mppi}
\begin{equation}\label{eq:mppi}
    \mathbf{W}_{out} = \left (\mathbf{X}^\intercal \mathbf{X} + \beta \mathbb{I} \right ) ^{-1} \mathbf{X}^\intercal \mathbf{Y}
\end{equation}

\noindent and using singular value decomposition (SVD) in \eqref{eq:mppi_svd}
\begin{equation}\label{eq:mppi_svd}
    \mathbf{W}_{out} = \left( \mathbf{V} \frac{\bm{\Sigma}}{\bm{\Sigma}{\odot}\bm{\Sigma} + \beta \mathbb{I}} \mathbf{U}^\intercal \right) \mathbf{Y}
\end{equation}

\noindent where $\beta$ is a regularization term, $\mathbb{I}$ is the identity matrix, $\odot$ is the Hadamard product, and $\mathbf{X} = \mathbf{U}\bm{\Sigma}\mathbf{V}^\intercal$. The SVD solution of the Moore-Penrose pseudoinverse gives a more accurate result but comes at the cost of higher computational complexity.

To maintain reservoir stability, ESNs need to satisfy the \emph{echo state property} (ESP) \cite{jaeger2001echo, gallicchio_echo_2017} as stated by \eqref{eq:echo_state_property}. 
\begin{equation}\label{eq:echo_state_property}
    \max_{\substack{1\le l \le N_L\\1\le k \le N_R}} {\left| \textup{eig}_k \left( (1-a^{(l)})\mathbf{\mathbb{I}}+a^{(l)}\hat{\mathbf{W}}_{res}^{(l)} \right) \right| < 1}
\end{equation}

\noindent The function $\textup{eig}_k$ gives the $k$\textsuperscript{th} eigenvalue of its matrix argument and $|{\cdot}|$ gives the modulus of its complex scalar argument. The maximum eigenvalue modulus is referred to as the spectral radius and must be less than unity (`1') in order for initial conditions of each reservoir to be washed out asymptotically. A hyperparameter $\hat{\rho}$ is substituted for unity to allow for reservoir tuning for a given task. Each $\hat{\mathbf{W}}_{res}^{(l)}$ is drawn from a uniform distribution and scaled such that the ESP is satisfied.

The remaining untrained weight matrices are set using one of the two methods. First, a matrix can be drawn from a uniform distribution and scaled to have a specified $L_2$ (Frobenius) norm, i.e.
\begin{align*}
    \mathbf{W}_{in}&= \frac{\mathbf{W}^\prime_{in}}{\norm{\mathbf{W}^\prime_{in}}_2}  \pi_{in},~
    \mathbf{W}_{res}^{(l)}= \frac{\mathbf{W}_{res}^{\prime(l)}}{\norm{\mathbf{W}_{res}^{\prime(l)}}_2}  \pi_{res}
\end{align*}

\noindent where $\pi_{in}$ and $\pi_{res}$ are hyperparameters.
Second, Glorot (Xavier) initialization \cite{glorot2010understanding} can be utilized without incurring further hyperparameters. The method initializes weights such that the activation (output) variance of consecutive fully-connected layers is the same. Formally, weights are drawn from a normal distribution with zero mean and a standard deviation of $\sqrt{{2} / {(n_{in} + n_{out})}}$, where $n_{in}$ and $n_{out}$ are the number of inputs and outputs of a layer, respectively.
All weight matrices are drawn with a sparsity parameter which gives the probability that each weight is nullified. Specifically, $s_{in}$ determines sparsity for $\mathbf{W}_{in}$, $\hat{s}_{res}$ for each $\hat{\mathbf{W}}_{res}^{(l)}$, and $s_{res}$ for each $\mathbf{W}^{(l)}_{res}$.

Furthermore, an unsupervised intrinsic plasticity (IP) learning rule is employed. The rule, originally proposed by \citeauthor{schrauwen_improving_2008}, introduces gain and bias terms to the nonlinearities of reservoir neurons, i.e. $\tanh(x)$ is substituted with $\tanh(g x + b)$ where $g$ is the gain and $b$ is the bias. Iterative application of the rule minimizes the Kullback-Leibler (KL) divergence between the empirical output distribution (as driven by $\mathcal{U}$) and a target Gaussian distribution \cite{schrauwen_improving_2008}. The update rules for the $i$\textsuperscript{th} neuron are given by \eqref{eq:ip_b} and \eqref{eq:ip_g}
\begin{align}
    \begin{split}\label{eq:ip_b}
        \Delta b^{(l)}_i(t{+}1) & = -\frac{\eta}{\sigma^2} \left( -\mu + \tilde{x}^{(l)}_i(t) \left( 2 \sigma^2 + 1 - \left(\tilde{x}^{(l)}_i(t) \right)^2 + \mu \tilde{x}^{(l)}_i(t) \right) \right)
    \end{split}\\
    \begin{split}\label{eq:ip_g}
        \Delta g^{(l)}_i(t{+}1) & = \frac{\eta}{g^{(l)}_i(t)} + \Delta b^{(l)}_i(t{+}1)~ x^{(l)}_i(t)
    \end{split}
\end{align}

\noindent where $\tilde{x}_i$ is given by \eqref{eq:layer_output_a} and $x_i$ is given by \eqref{eq:layer_output_b}. The hyperparameter $\eta$ is the the learning rate, and $\sigma$ and $\mu$ are the standard deviation and mean of a target Gaussian distribution, respectively. In a pre-training phase, the learned parameters are each initialized as $b^{(l)}_i(t)=0$ and $g^{(l)}_i(t)=1$ and are updated iteratively in a layer-wise fashion.

\subsection{Particle Swarm Optimization}
Instances of the \emph{Mod-DeepESN} network may achieve satisfactory forecasting performance using empirical guesses of hyperparameters, however, a more sophisticated optimization approach will further improve the performance.
We thus propose black-box optimization of the \emph{Mod-DeepESN} hyperparameters using particle swarm optimization (PSO) \cite{kennedy1995particle}. In PSO, a population of particles is instantiated in the search space of an optimization problem. This space contains the possible values of continuous or discrete hyperparameters and the particles move around the space based on external fitness, or cost, signals. The communication network topology employed dictates the social behavior, or dynamics, of the swarm. Here, we utilize a star topology in which each particle is attracted to the globally best-performing particle.

Formally, each particle is a candidate solution of $N_H$ hyperparameters with position $\mathbf{p}_i(t) \in \mathbb{R}^{N_H}$ and velocity $\mathbf{v}_i(t) \in \mathbb{R}^{N_H}$. The trivial position update of each particle is given by \eqref{eq:particle_position_update} while the velocity update is given by \eqref{eq:particle_velocity_update}.
\begin{align}
    \begin{split}\label{eq:particle_position_update}
        \mathbf{p}_i(t+1)    =~& \mathbf{p}_i(t) + \mathbf{v}_i(t+1)
    \end{split}\\
    \begin{split}\label{eq:particle_velocity_update}
        \mathbf{v}_{i}(t+1) =~& w \mathbf{v}_{i}(t) + \varphi_1 \mathbf{U}_{1}(t)\left(\mathbf{\hat{b}}_{i}(t) - \mathbf{p}_{i}(t)\right) \\
                     & + \varphi_2 \mathbf{U}_2(t)\left(\mathbf{\hat{b}}^*_i(t) - \mathbf{p}_i(t)\right)
    \end{split}
\end{align}

\noindent The matrices $\mathbf{U}_1(t),\mathbf{U}_2(t) \in \mathbb{R}^{N_H \times N_H}$ are populated by values uniformly drawn from the interval $[0, 1)$ at each timestep. The best position found by a particle is the vector $\mathbf{\hat{b}}_i(t) \in \mathbb{R}^{N_H}$ while the best solution found in the neighborhood of a particle is the vector $\mathbf{\hat{b}}^*_i(t) \in \mathbb{R}^{N_H}$. With a star communication topology, $\mathbf{\hat{b}}^*_i(t)$ is the best position found globally. The velocity update comprises three parameters which influence a particle's dynamics: $w$ is the inertia weight, $\varphi_1$ is the cognitive acceleration (how much a particle should follow its personal best), and $\varphi_2$ is the social acceleration (how much a particle should follow the swarm's global best). All hyperparameters are considered during the optimization process, except for $\beta$, which is swept after $\mathbf{X}$ has been computed, exploiting the fact that all the weight matrices but $\mathbf{W}_{out}$ are fixed.

\subsection{Neural Mapping}
RC networks have strong underpinnings in neural processing. Recent studies have shown that the distribution of the complex representation layer and the linear construction layer in the reservoir is similar to the one observed in the cerebellum. The model with granule layer (representation layer) and the synapses between granule and Purkinje cells (linear readout layer)~\cite{cereb_2007} is used to study computationally useful case studies such as vestibulo-ocular reflex adaptation~\cite{dean2002decorrelation}. The Purkinje cells are trained from the random, distributed input signals of the granule cells' parallel dendritic connections \cite{granule_cells}. We also employ intrinsic plasticity, as in \cite{schrauwen_improving_2008}, to modulate neuronal activations to follow a known distribution. Whereas a biological neuron's electrical properties are modified, a reservoir neuron's gain and bias are adjusted. There are arguments that identifying the powerful computational paradigm, edge-of-chaos, for the biological counterparts of the reservoir are yet to be understood. However, infusing hybrid plasticity mechanisms such as short-term plasticity or long-term synaptic plasticity can help enhance the computational performance. It is interesting to note that the reservoir computational models (both spiking and non-spiking) seem to have a boost in their performance from embedding intrinsic plasticity, akin to the biological models \cite{soures2017deep,soures2017device}. This convergence with the biological counterparts vastly improves our understanding of building spatiotemporal processing, though one should take a parsimonious approach with correlations.


\section{Measuring Reservoir Goodness}

Various metrics have been proposed to understand reservoir performance \cite{jaeger2002short, ozturk_esn_analysis_2007, gibbons_unifying_2010, gallicchioLocalLyapunovExponents2018, lymburnConsistencyEchostateNetworks2019}. In this work, we quantify reservoir goodness by measuring the stability of reservoir dynamics as well as evaluating the forecasting proficiency of networks trained on synthetic and real-world tasks.

\subsection{Separation Ratio Graphs}

Separation ratio graphs \cite{gibbons_unifying_2010} are considered for determining the fitness of \emph{Mod-DeepESN} state separability. The method compares the separation of inputs with the separation of outputs for a given input and output, e.g. the input and output of a single reservoir or the input and output of a set of reservoirs. The metric assumes that inputs that appear similar should have a comparable degree of similarity with outputs. That is,
\begin{equation}\label{eq:sep_ratio}
    \underbrace{\overbrace{\frac{
        \norm{ \mathbf{x}^{(l_2)}_i(t) - \mathbf{x}^{(l_2)}_j(t) }_2
    }{
        \norm{ \mathbf{i}^{(l_1)}_i(t) - \mathbf{i}^{(l_1)}_j(t) }_2
    }}^{\textup{Output Separation}}}_{\textup{Input Separation}}
    \approx 1, ~l_2 > l_1
\end{equation}

\noindent where $\norm{\cdot}_2$ gives the Euclidean distance between inputs (or outputs) $i$ and $j$. When the output separation is plotted as a function of the input separation, the relation should be close to identity, i.e. a linear regression trend line should yield a slope $m \approx 1$ and intercept $b \approx 0$. If the output separation $\gg$ the input separation, a reservoir (or network) is considered to be in the ``chaotic'' zone, whereas it is considered to be in the ``attractor'' zone if the output separation $\ll$ the input separation.

\begin{figure*}
    \centering
    \includegraphics[width=.8\linewidth]{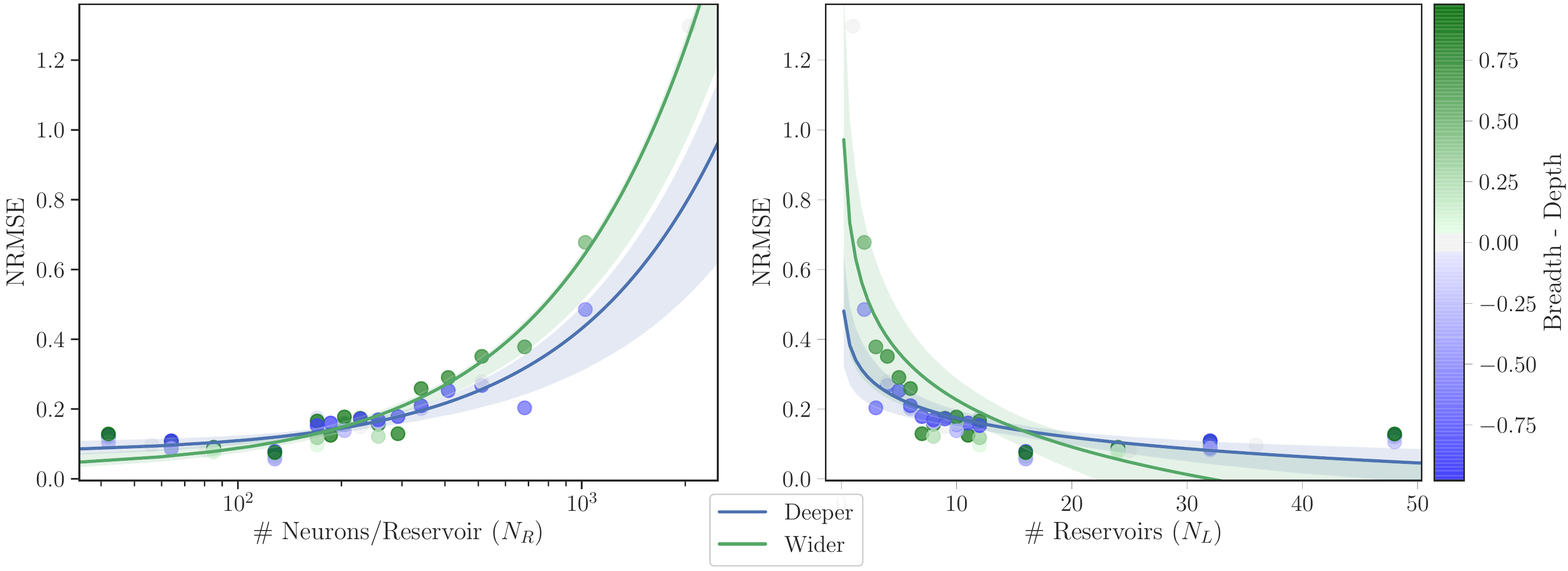}
    \caption{Reservoir neuronal budgeting results for $N_N=2048$ neurons for the Mackey Glass task. Color corresponds to $\frac{N_{LB} - N_{LD}}{N_L}$. \textit{Left}: NRMSE as a function of $N_R$ with a linear trend line. \textit{Right}: NRMSE as a function of $N_L$ with a linear trend line of NRMSE as a function of $\log(N_L)$.}
    \label{fig:mackey_budget}
\end{figure*}

\subsection{Lyapunov Exponent}

The Lyapunov exponent (LE) offers a quantitative measure of the exponential growth (decay) rate of infinitesimal perturbations within a dynamical system \cite{lyapunov1992general, Eckmann2004, gallicchioLocalLyapunovExponents2018}. An $N$-dimensional system has $N$ LEs which describe the evolution along each dimension of its state space. The maximum LE (MLE) is a good indication of a system's stability as it commands the rate of contraction or expansion in the system state space. If the MLE of a system is below 0, a system is said to exhibit ``stable'' dynamics, whereas an MLE above 0 describes a system with ``chaotic'' dynamics. An MLE of 0 is commonly referred to as the ``edge of chaos.'' Local MLEs \cite{gallicchioLocalLyapunovExponents2018} are considered in this work as they are more useful in practical experiments and can be estimated by driving a network by a real input signal (e.g. a time series). The MLE, denoted $\lambda_{max}$, of a \emph{Mod-DeepESN} instance can be computed for a given set of input sequences by \eqref{eq:lambda_max}
\begin{multline}\label{eq:lambda_max}
    \lambda_{max} = \max_{\substack{1 \le l \le N_L\\1 \le k \le N_R}}
    \frac{1}{N_S N_t} \sum_{i=1}^{N_S} \sum_{t=1}^{N_t}
    \ln \left(\left|\text{eig}_k\left( \left(1 - a ^ {(l)}\right) \mathbb{I}
    \right.\right.\right.\\[-2.5ex]
    \left.\left.\left.
    +~ a ^ {(l)} \mathbf{D}_i^{(l)}(t) \hat{\mathbf{W}}^{(l)} \right ) \right| \right)
\end{multline}

\noindent where for the $i$\textsuperscript{th} sequence the diagonal matrix $\mathbf{D}_i^{(l)}(t)$ is given by \eqref{eq:the_D}.
\begin{equation}\label{eq:the_D}
    \mathbf{D}_i^{(l)}(t) =
    \begin{bmatrix}
        1 - \left(\tilde{x}^{(l)}_1(t)\right)^2 & 0 & \hdots & 0 \\
        0 & 1 - \left(\tilde{x}^{(l)}_{2}(t)\right)^2 & \hdots & 0 \\
        \vdots & \vdots & \ddots & \vdots \\
        0 & 0 & \hdots & 1 - \left(\tilde{x}^{(l)}_{N_R}(t)\right)^2
    \end{bmatrix}
\end{equation}

\subsection{Task Performance Metrics}

The more straightforward way of measuring reservoir goodness is by evaluating the performance of a network on a task. For scalar-valued time series forecasting tasks, we consider the following three metrics to quantify network error: root-mean-square error (RMSE) \eqref{eq:rmse}, normalized RMSE (NRMSE) \eqref{eq:nrmse}, and mean absolute percentage error (MAPE) \eqref{eq:mape}.
\begin{align}
    \begin{split}\label{eq:rmse}
        \textup{RMSE} &= \sqrt{\frac{1}{N_S N_t} \sum_{i=1}^{N_S}\sum_{t=1}^{N_t} \left(\mathbf{y}(t) - \hat{\mathbf{y}}(t)\right)^2}
    \end{split}\\
    \begin{split}\label{eq:nrmse}
        \textup{NRMSE} &= \sqrt{\frac{
            \sum_{i=1}^{N_S}\sum_{t=1}^{N_t} \left(\mathbf{y}(t) - \hat{\mathbf{y}}(t)\right)^2
        }{
            \sum_{i=1}^{N_S}\sum_{t=1}^{N_t} \left(\mathbf{y}(t) - \bar{\mathbf{y}}\right)^2
        }}
    \end{split}\\
    \begin{split}\label{eq:mape}
        \textup{MAPE} &= \frac{100\%}{N_S N_t} \sum_{i=1}^{N_S}\sum_{t=1}^{N_t} \frac{|\mathbf{y}(t) - \hat{\mathbf{y}}(t)|}{\mathbf{y}(t)}
    \end{split}
\end{align}

\noindent The vector $\hat{\mathbf{y}}(t)$ is the predicted time series value at timestep $t$, $\bar{\mathbf{y}}$ is the average value of the ground truth (time series) over the $N_t$ timesteps, $\mathbf{y}(t) \in \mathbf{Y}$, and $\hat{\mathbf{y}}(t), \mathbf{y}(t), \bar{\mathbf{y}} \in \mathbb{R} ^ {N_Y}$.

Proposed in \cite{bay2009evaluation} and adapted for polyphonic music tasks in \cite{boulanger2012modeling}, frame-level accuracy (FL-ACC) rewards only true positives across every timestep of all samples as shown in \eqref{eq:fl-acc}.
\begin{equation}\label{eq:fl-acc}
    \textup{FL-ACC} = \frac{
        \sum_{i=1}^{N_S}\sum^{N_t}_{t=1} \textup{TP}_i(t)
    }{
        \sum_{i=1}^{N_S}\sum^{N_t}_{t=1} \textup{TP}_i(t) + \textup{FP}_i(t) + \textup{FN}_i(t)
    }
\end{equation}

\noindent The subscript $i$ of each of the \{true, false\} positives (\{T,F\}P) and false negatives (FN) denotes the corresponding quantity for the $i$\textsuperscript{th} sequence. Note that FL-ACC is analogous to the intersection-over-union (IoU) metric, also known as the Jaccard index \cite{jaccard1912}, which is commonly used to assess the performance of image segmentation models.

For forecasting tasks, NRMSE is considered as its value is not dependent on the scale of the model nor the data, thus allowing for a more accurate comparison with results reported in the literature. FL-ACC is utilized for the polyphonic forecasting task as the true negative (TN) rate is not a good indicator of performance, as well as for consistent comparison of results.


\begin{figure*}
    \centering
    \includegraphics[width=.8\linewidth]{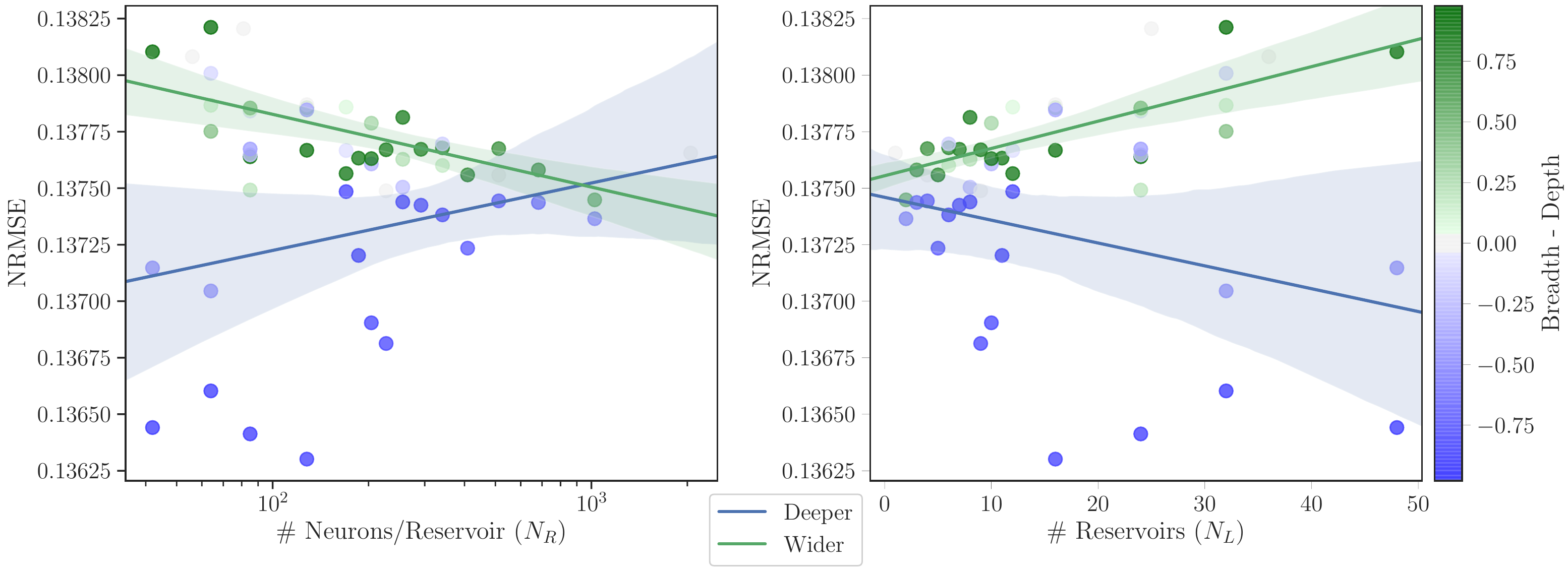}
    \caption{Reservoir neuronal budgeting results for $N_N=2048$ neurons for the Melbourne, Australia, minimum temperature forecasting task. Color corresponds to $\frac{N_{LB} - N_{LD}}{N_L}$. \textit{Left}: NRMSE as a function of $N_R$ with a linear trend line of NRMSE as a function of $\log(N_R)$. \textit{Right}: NRMSE as a function of $N_L$ with a linear trend line.}
    \label{fig:melbourne_budget}
\end{figure*}

\section{Experiments}

We analyze the \emph{Mod-DeepESN} architecture on diverse time series forecasting tasks that exhibit multi-scale and nonlinear dynamics: the chaotic Mackey-Glass series, a daily minimum temperature series \cite{temp_data}, and a set of polyphonic music series \cite{poliner2006,boulanger2012modeling}.

\subsection{Practical Details}
For each task considered in this work, we run PSO to produce a candidate set of hyperparameters that offer the best performance on the appropriate validation set of data. Thereafter, these parameterizations of \emph{Mod-DeepESN} are evaluated on the test set for the specific task. All numerical results are averaged over 10 runs. We run PSO for 100 iterations with 50 particles and set its parameters as follows: $\varphi_1 = 0.5$, $\varphi_2 = 0.3$, $w = 0.9$.

During training, $\beta$ is swept from the set $\{0\} \cup \{10 ^ {-n} ~|~ n \in [1 .. 8] \}$,
exploiting the fact that $\mathbf{X}$ need only be computed once. Ridge regression is carried out using SVD according to \eqref{eq:mppi_svd} for increased numerical stability. Only the score produced by the best-performing $\beta$ is considered during the PSO update.

We only consider dense grid topologies in these experiments, which are referred to as \textit{Wide} and \textit{Layered} in \cite{Mod-DeepESN}. This allows all networks to be described in terms of breadth and depth, reducing the complexity of neuronal partitioning and other analyses. Additionally, the leaky rate $\alpha$ is kept constant across all the reservoirs, i.e. $\alpha = a^{(l)}~\forall l \in [1..N_L].$

\subsubsection{Model Implementation}
The \textit{Mod-DeepESN} architecture and its utilities are implemented in Python using several open-source libraries. \verb|TensorFlow| \cite{tensorflow} and \verb|Keras| \cite{keras} are used for matrix/tensor operations, managing network graphs, and unrolling RNNs. The PSO implementation extends the base optimizers provided by the \verb|PySwarms| \cite{pyswarms} library, and the \verb|pandas| \cite{pandas}, \verb|NumPy| \cite{numpy}, and \verb|SciPy| \cite{scipy} libraries are employed throughout the codebase.

\subsection{Neuronal Partitioning}
An interesting question in reservoir models is to determine the optimal size and connectivity of/within reservoirs. Is a large reservoir as effective as a multitude of small reservoirs, and should these small reservoirs extend deeper or wider? To address this matter, we propose \textit{neuronal partitioning} to explore a space of grid topologies. Given a budget of $N_N$ neurons network-wide, we evaluate the task performance for a depth $N_{LD}$ and breadth $N_{LB}$ where $N_L = N_{LD} \times N_{LB}$ and $N_R = \lfloor N_N / N_L \rfloor$. This experiment is performed for the Mackey Glass and minimum temperature forecasting tasks. The prohibitive size of the polyphonic music data prevents us from running such an experiment. The values of $N_{LD}$ and $N_{LB}$ selected for each experiment are the integer factors of $N_L \in [1 .. 12]$ and of $N_L \in \{16, 24, 25, 32, 36, 48, 49, 64\}$.

\begin{figure}[b]
    \centering
    \includegraphics[width=.8\linewidth]{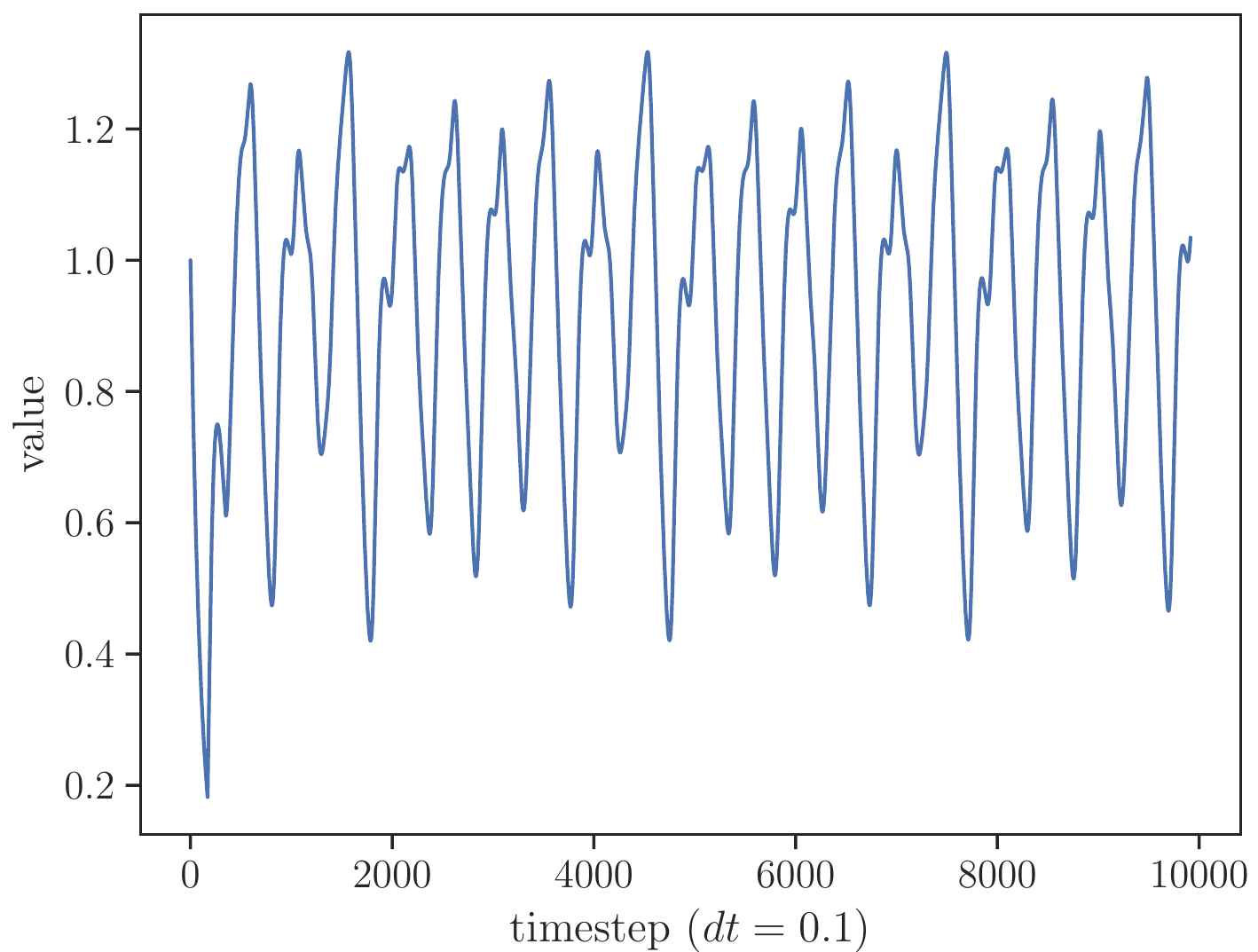}
    \caption{The Mackey Glass chaotic time series \cite{MackeyGlass} computed over 10,000 timesteps of duration $dt = 0.1$.}
    \label{fig:mackey_glass}
\end{figure}

\begin{figure*}
    \centering
    \begin{subfigure}{.333\linewidth}
      \centering
      \includegraphics[width=\linewidth]{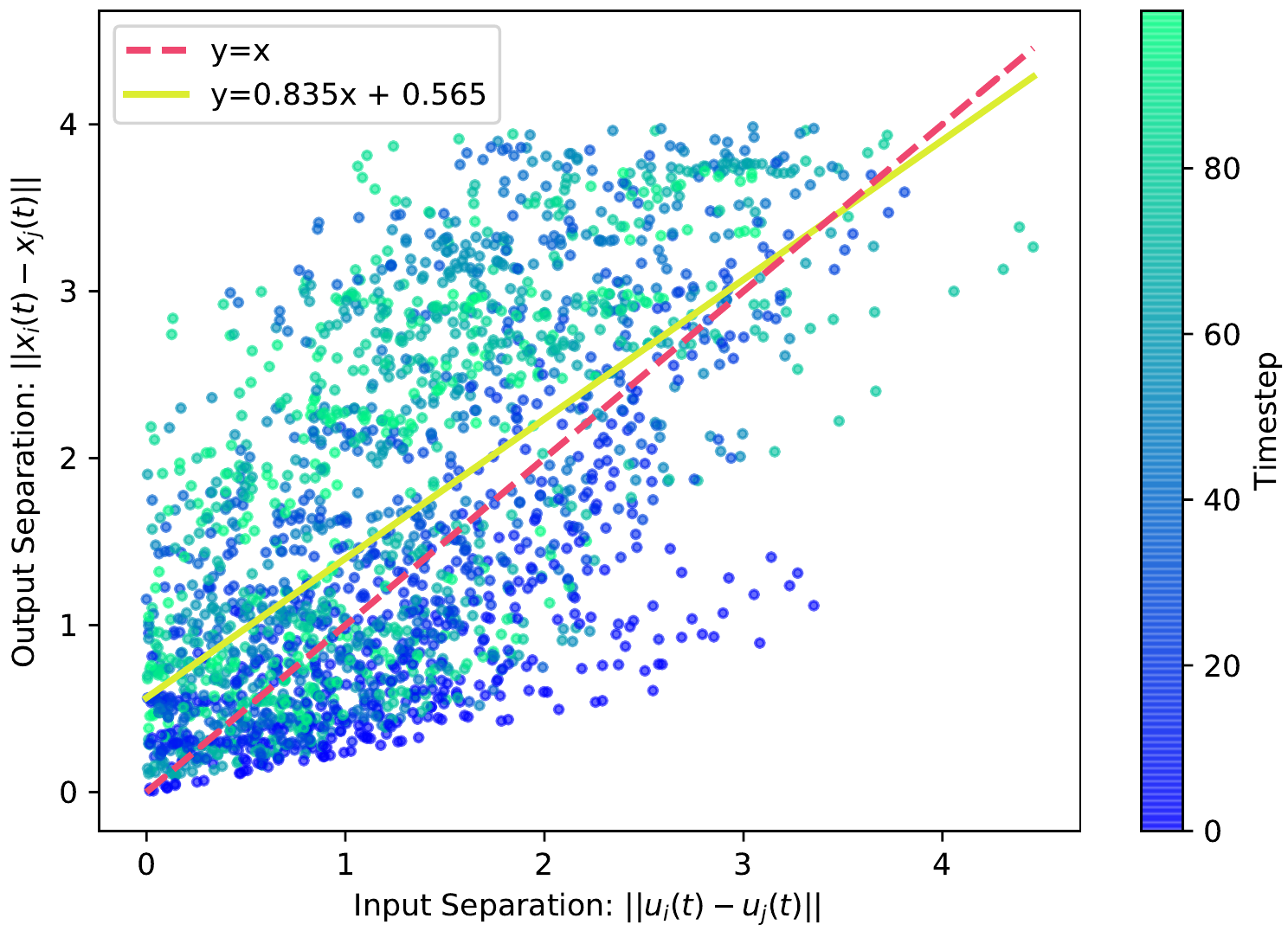}
      \caption{}
      \label{fig:sr_rank1}
    \end{subfigure}%
    \begin{subfigure}{.333\linewidth}
      \centering
      \includegraphics[width=\linewidth]{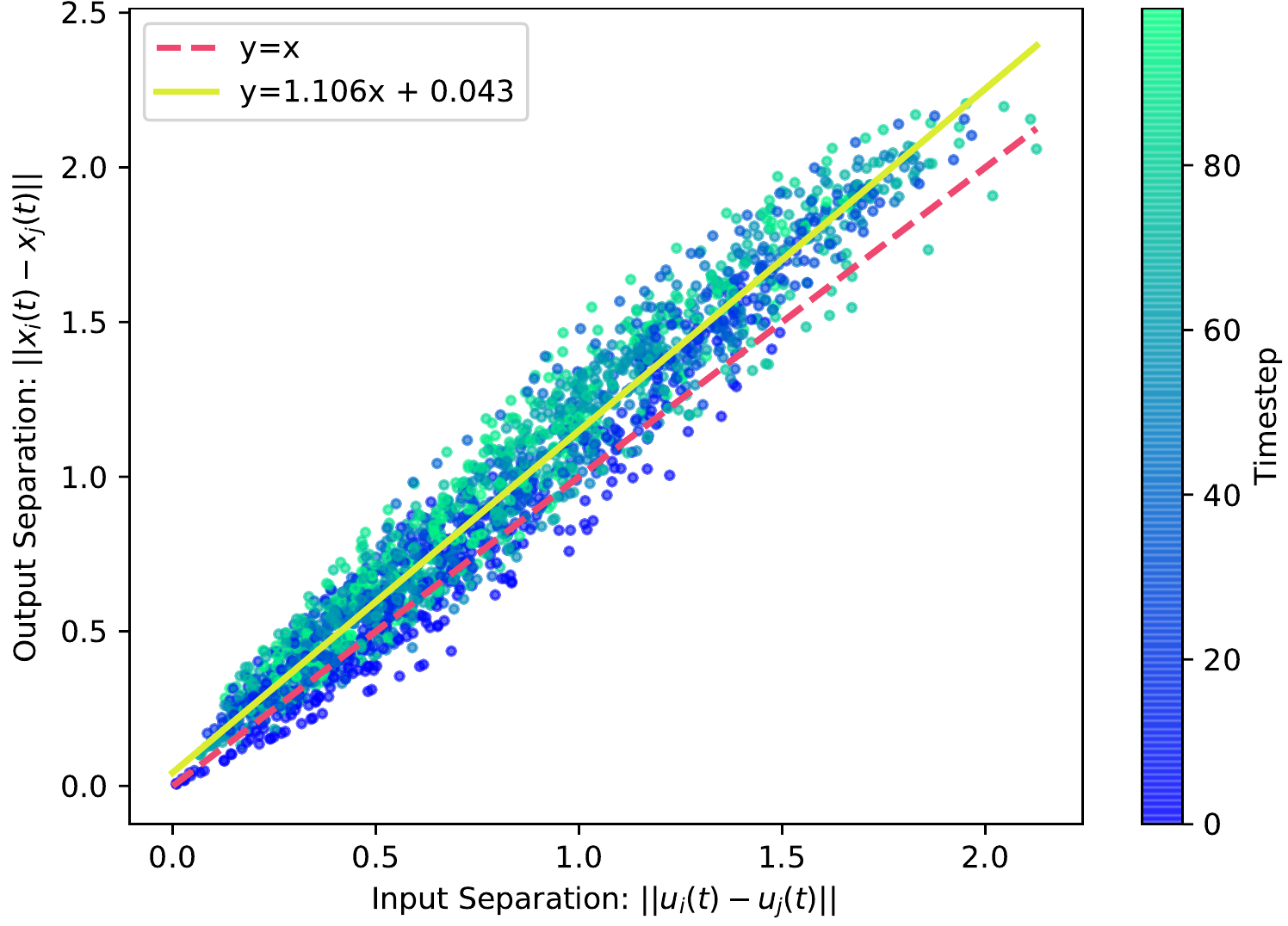}
      \caption{}
      \label{fig:sr_rank2}
    \end{subfigure}%
    \begin{subfigure}{.333\linewidth}
      \centering
      \includegraphics[width=\linewidth]{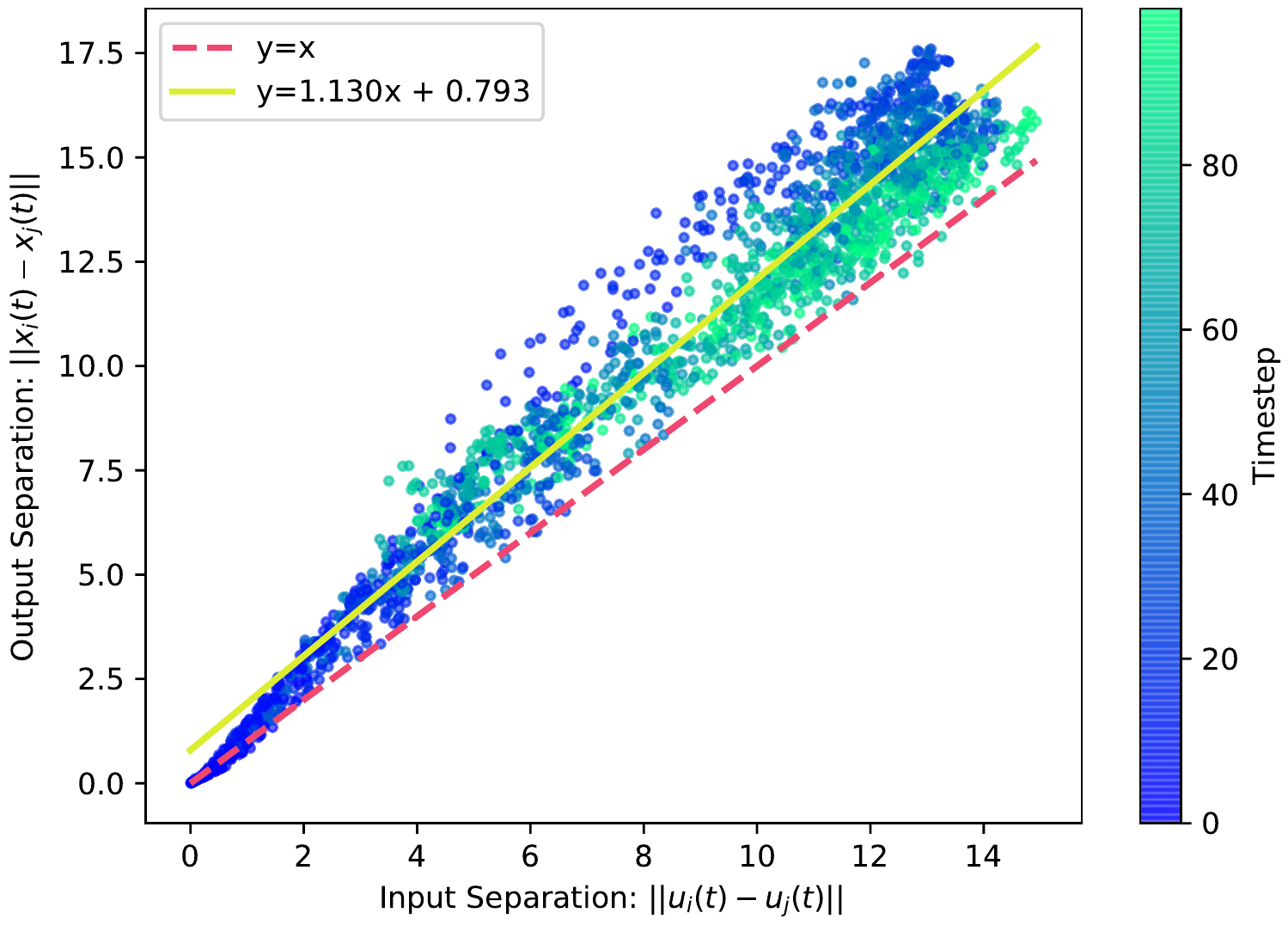}
      \caption{}
      \label{fig:sr_rank_last}
    \end{subfigure}%
    \vspace{-3.5mm}
    \caption{Separation ratio plots for various \textit{Mod-DeepESN} instances for the Melbourne forecasting task. (a) Separation ratio plot for the best-performing model. (b) Separation ratio plot for the second-best-performing model. (c) Separation ratio plot for the worst-performing model.}
    \label{fig:sep_ratios}
\end{figure*}

\subsection{Mackey Glass}

Mackey Glass \cite{MackeyGlass} is a classical chaotic time series benchmark for evaluating the forecasting capacity of dynamical systems \cite{jaeger2001echo,yusoff_modeling_2016,aceitunoTailoringArtificialNeural2017,maDeepESNMultipleProjectionencoding2017}. The series is generated from a nonlinear time delay differential equation using the fourth-order Runge-Kutta method (RK4) and is given by \eqref{eq:mackey}.
\begin{equation}\label{eq:mackey}
    \frac{dx}{dt} =
    \beta \frac{x(t - \tau)}{1 + x(t - \tau) ^ n} - \gamma x(t)
\end{equation}

\noindent During generation, we set $\tau$=17, $\beta$=0.2, $\gamma$=0.1, and $n$=10 with a time resolution ($dt$) of 0.1 to compare with methods evaluated in \cite{maDeepESNMultipleProjectionencoding2017}. 10,000 samples are split into 6,400 training samples 1,600 validation samples, and 2,000 testing samples for 84 timestep-ahead forecasting, i.e. given $\mathbf{u}(t)$, predict $\mathbf{y}(t) = \mathbf{u}(t+84)$. To reduce the influence of transients, the first 100 timesteps of the training set are used as a washout period.

Table \ref{tab:mackey_results} contains the best forecasting results for \emph{Mod-DeepESN} as well as those reported in \cite{maDeepESNMultipleProjectionencoding2017}. The \emph{Mod-DeepESN} framework falls slightly short of the performance of Deep-ESN but outperforms the other baselines in terms of (N)RMSE. MAPE exhibits several biases, such as punishing negative errors more than positive, which may be the reason for this discrepancy.

\begin{table}[b]
  \centering
  \caption{Mackey-Glass Time Series 84-Step Ahead Prediction Results. All errors are reported in thousandths.}
  \label{tab:mackey_results}
  \ra{1.1}
  \begin{tabular}{@{}*{5}{c}@{}}
    \toprule
    Network & $N_L$ & RMSE & NRMSE & MAPE\\
    \midrule
    Vanilla ESN  \cite{jaeger_optimization_2007}
    & 1 & 43.7 & 201 & 7.03  \\
    $\phi$-ESN \cite{phi_2011}
    & 2 & 8.60 & 39.6 & 1.00 \\
    R$^2$SP    \cite{r2sp_2013}
    & 2 & 27.2 & 125 & 1.00 \\
    MESM	     \cite{mesm_2017}
    & 7 & 12.7 & 58.6 & 1.91 \\
    Deep-ESN   \cite{maDeepESNMultipleProjectionencoding2017}
    & 2 & \textbf{1.12} & \textbf{5.17} & \textbf{.151} \\
    \textit{Mod-DeepESN} & 3 & {7.22} & {27.5} & {5.55} \\
    \bottomrule
  \end{tabular}
\end{table}

\subsubsection{Neuronal Partitioning}
Neural partitioning is run for the Mackey Glass task with results reported in Figure \ref{fig:mackey_budget}. It is apparent that smaller values of $N_R$ and larger values of $N_L$ yield the lowest NRMSE; an agglomeration of small reservoirs outperform a single large reservoir for this task. While marginal, broader topologies outperform deeper for this task.

\begin{table}[b]
  \centering
  \caption{Daily Minimum Temperature Series 1-Step Ahead Prediction Results. All errors are reported in thousandths.}
  \label{tab:temp_results}
  \ra{1.1}
  \begin{tabular}{@{}*{5}{c}@{}}
    \toprule
    Network & $N_L$ & RMSE & NRMSE & MAPE\\
    \midrule
    ESN 	     \cite{jaeger_optimization_2007}
        & 1 & 501  & 139  & 39.5  \\ 
    $\phi$-ESN \cite{phi_2011} 
        & 2 & 493  & 141  & 39.6  \\
    R$^2$SP    \cite{r2sp_2013}
        & 2 & 495  & 137  & 39.3  \\
    MESM	     \cite{mesm_2017}
        & 7 & 478  & 136  & 37.7  \\
    Deep-ESN   \cite{maDeepESNMultipleProjectionencoding2017} 
        & 2 & 473  & 135  & \textbf{37.0} \\
    \textit{Mod-DeepESN} & 4 & \textbf{459} & \textbf{132} & {37.1}\\
    \bottomrule
  \end{tabular}
\end{table}

\begin{figure}[t]
    \centering
    \includegraphics[width=.8\linewidth]{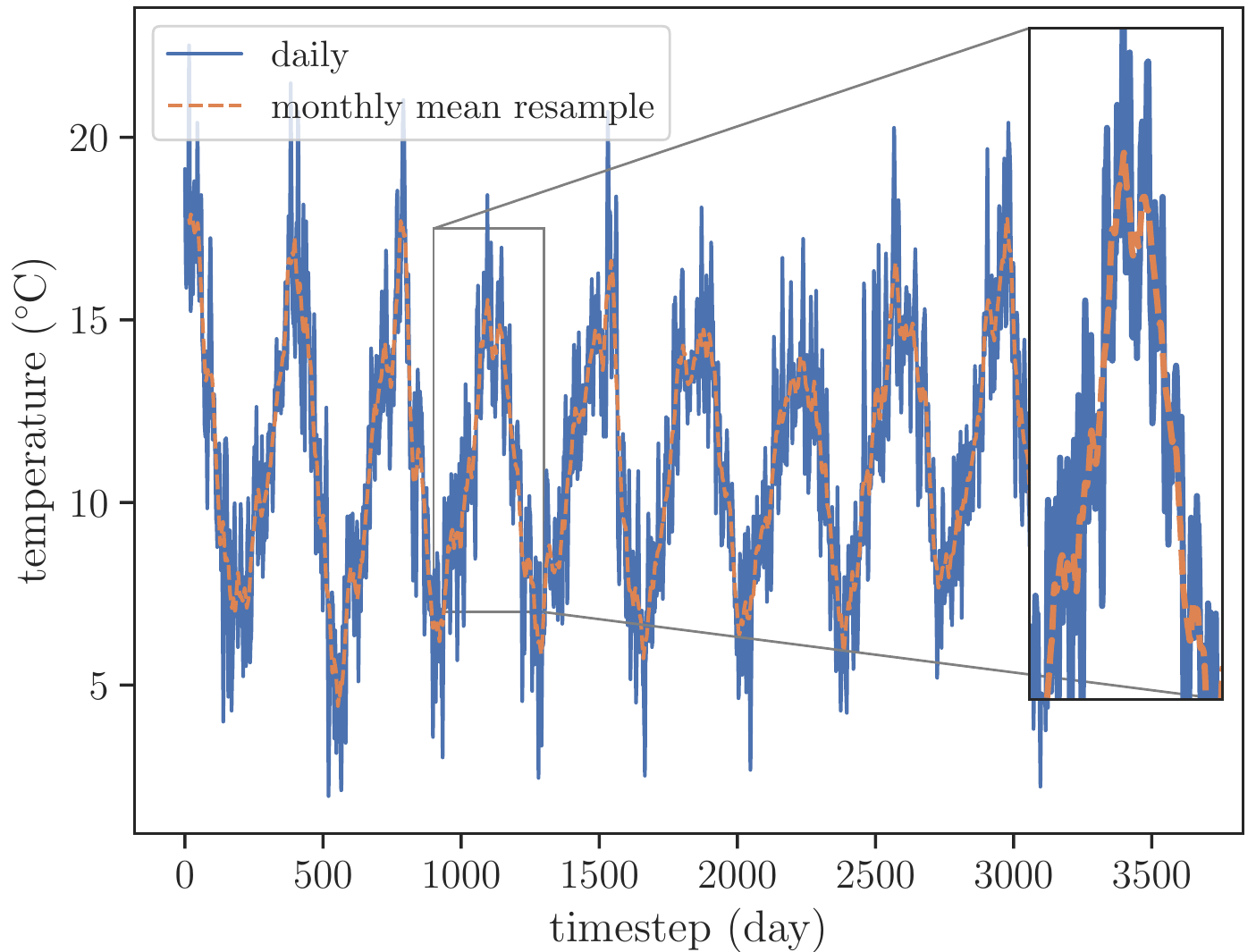}
    \caption{The Melbourne, Australia, daily minimum temperature time series \cite{temp_data}.}
    \label{fig:melbourne}
\end{figure}

\begin{figure*}
    \centering
    \begin{subfigure}{.05\linewidth}
      \centering
      \caption{}
    \end{subfigure}%
    \begin{subfigure}{.45\linewidth}
      \centering
      \includegraphics[width=.85\linewidth]{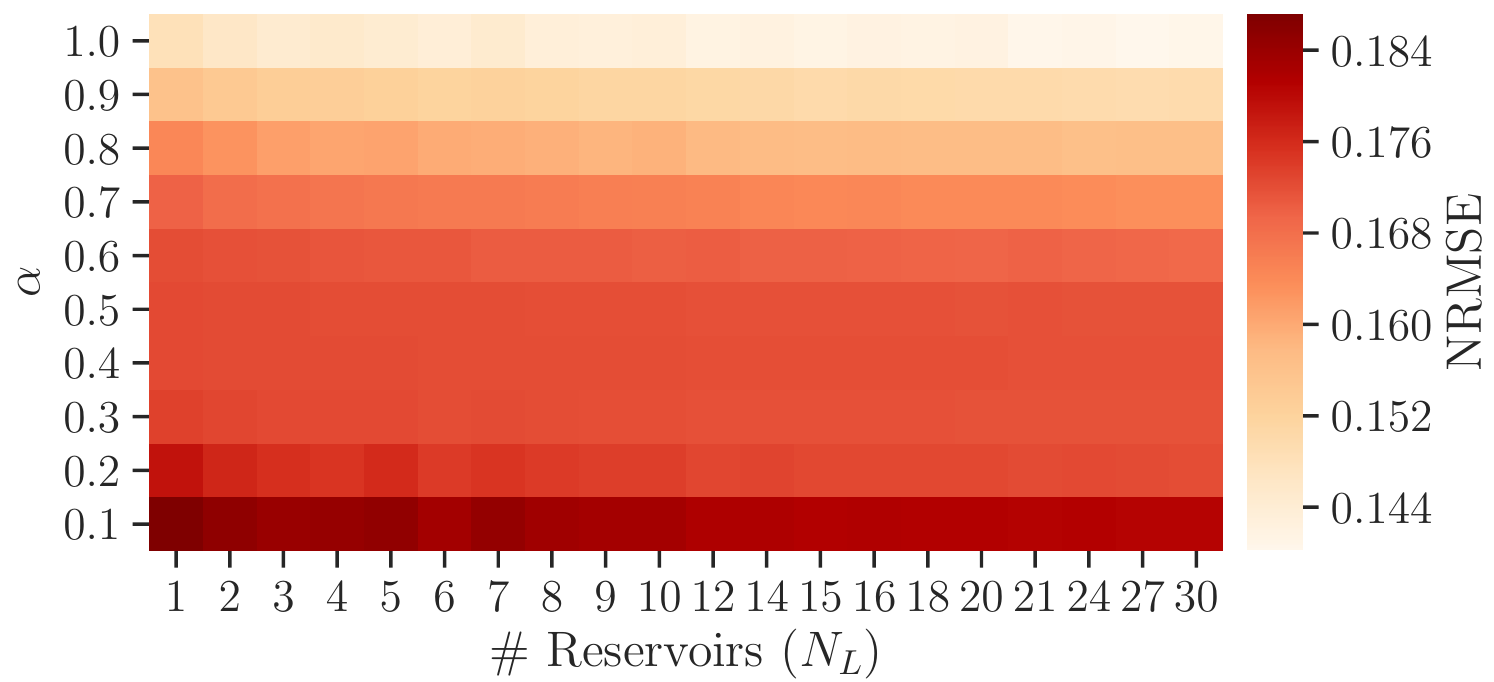}
      \label{fig:sweep_alpha_hm}
    \end{subfigure}%
    \begin{subfigure}{.05\linewidth}
      \centering
      \caption{}
    \end{subfigure}%
    \begin{subfigure}{.45\linewidth}
      \centering
      \includegraphics[width=.85\linewidth]{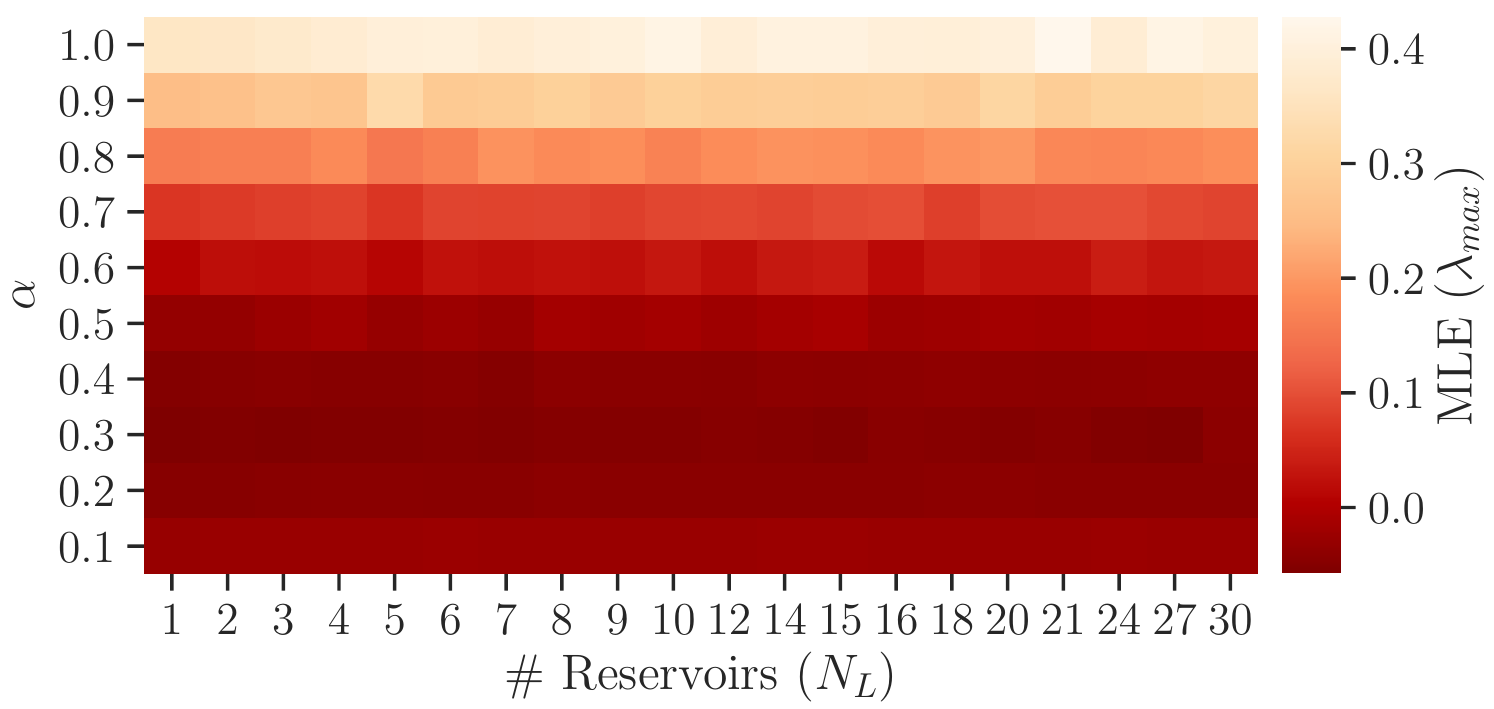}
      \label{fig:sweep_alpha_hm_lambda}
    \end{subfigure}\vskip\baselineskip\vspace{-3mm}
    \begin{subfigure}{.05\linewidth}
      \centering
      \caption{}
    \end{subfigure}%
    \begin{subfigure}{.45\linewidth}
      \centering
      \includegraphics[width=.85\linewidth]{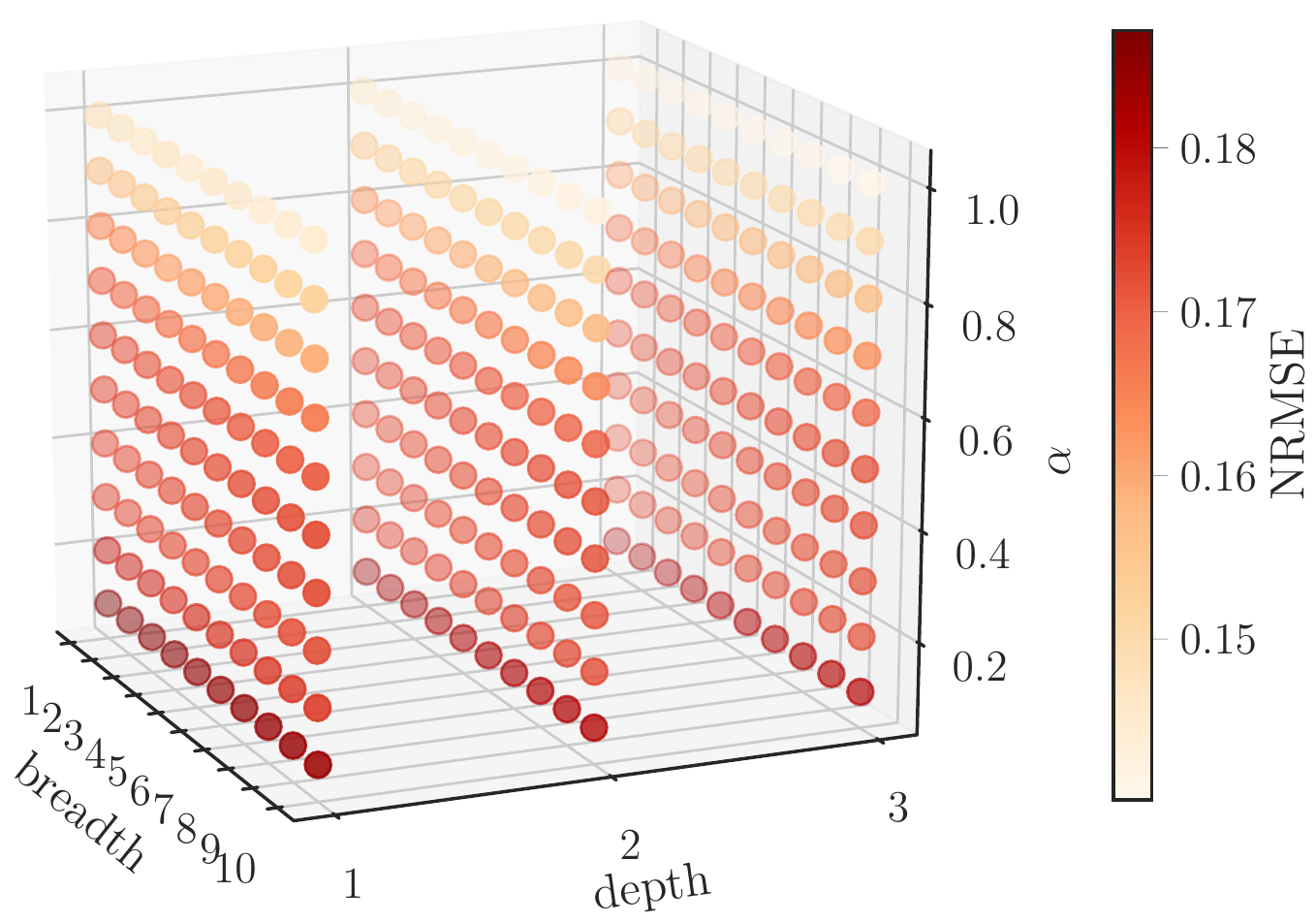}
      \label{fig:sweep_alpha_3d}
    \end{subfigure}%
    \begin{subfigure}{.05\linewidth}
      \centering
      \caption{}
    \end{subfigure}%
    \begin{subfigure}{.45\linewidth}
      \centering
      \includegraphics[width=.85\linewidth]{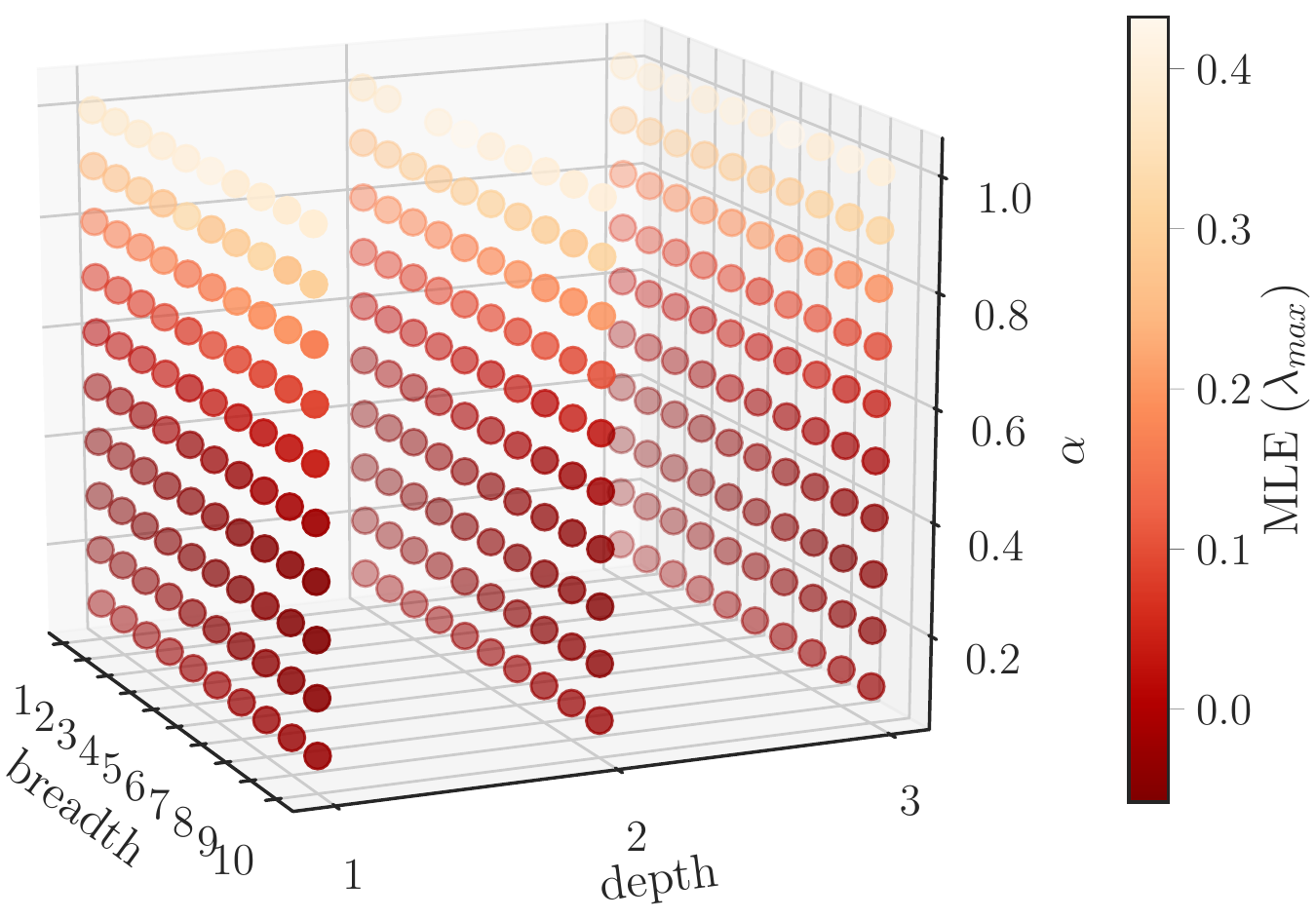}
      \label{fig:sweep_alpha_3d_lambda}
    \end{subfigure}%
    \caption{2D and 3D heatmaps of NRMSE and $\lambda_{max}$ as a function of $\alpha$ and $N_L$. Note that the color bar gradient is reversed for visualizing $\lambda_{max}$. (a) Impact of $N_L$ and $\alpha$ on NRMSE. (b) Impact of $N_L$ and $\alpha$ on $\lambda_{max}$. (c) Impact of reservoir breadth, depth, and $\alpha$ on NRMSE. (d) Impact of reservoir breadth, depth, and $\alpha$ on $\lambda_{max}$.}
    \label{fig:sweep_alpha}
\end{figure*}

\subsection{Melbourne, Australia, Daily Minimum Temperature Forecasting}
The Melbourne, Australia, daily minimum temperature forecasting series \cite{temp_data} is recorded from 1981-1990 and shown in Figure \ref{fig:melbourne}. In this task, the goal is to predict the next minimum temperature of the directly proceeding day in Melbourne, i.e. given $\mathbf{u}(t)$, predict $\mathbf{y}(t) = \mathbf{u}(t+1)$. The data is smoothed with a 5-step moving window average and split into 2,336 training samples, 584 validation samples, and 730 testing samples to compare with methods evaluated in \cite{maDeepESNMultipleProjectionencoding2017}. A washout period of 30 timesteps (days) is used to rid transients.

Table \ref{tab:temp_results} contains the best forecasting results for \emph{Mod-DeepESN} as well as those reported in \cite{maDeepESNMultipleProjectionencoding2017}. The \emph{Mod-DeepESN} framework outperforms all baselines in terms of (N)RMSE. This result is more interesting than that of Mackey Glass as the time series comprises real data as opposed to synthetic.
\begin{figure}[b]
    \centering
    \includegraphics[width=\linewidth]{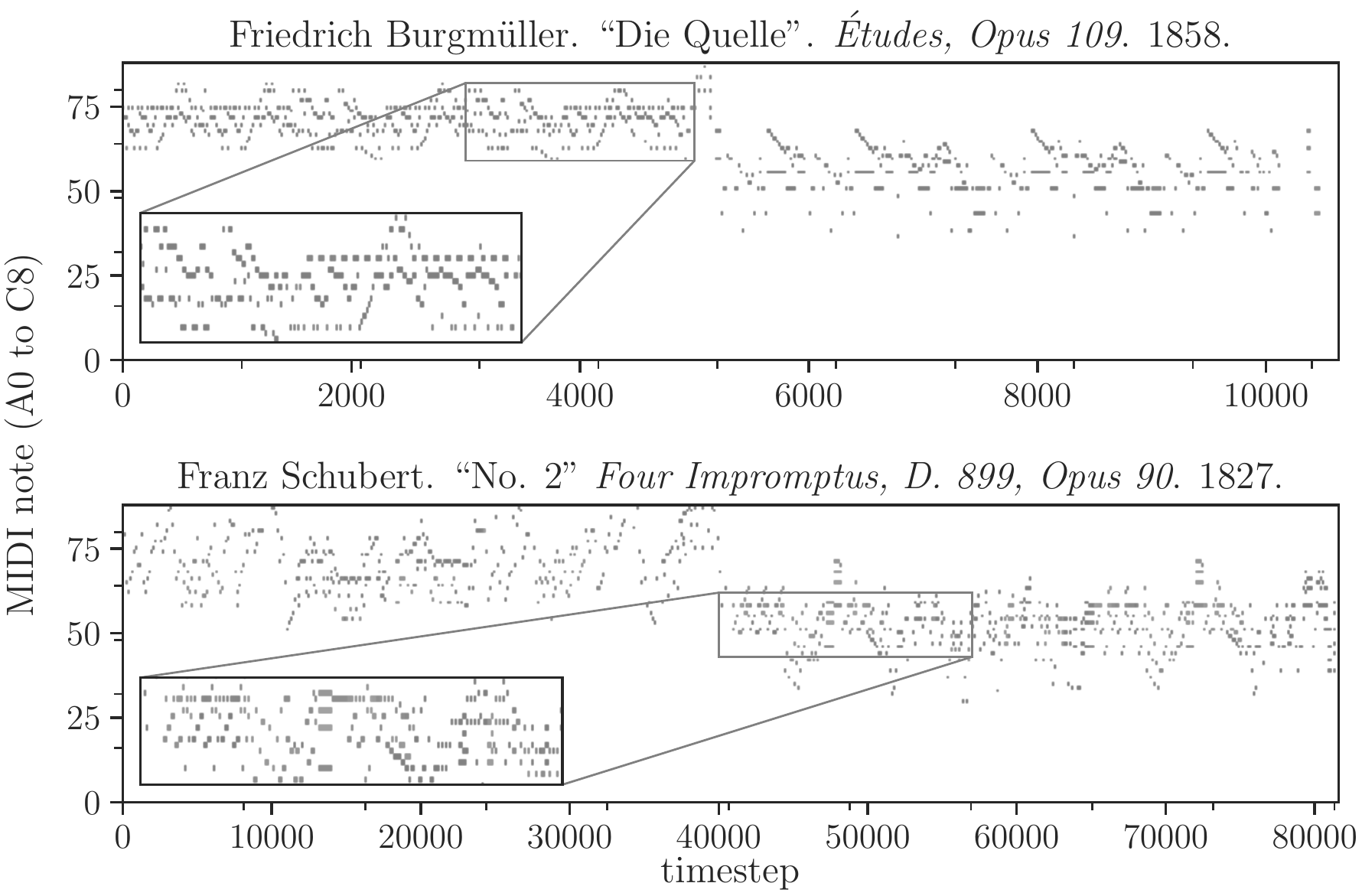}
    \caption{Two samples from the Piano-midi.de dataset with insets highlighting activity. Terminology: \textit{Opus}: number of a musical work indicating chronological order of production; \textit{D.}: \underline{D}eutsch Thematic Catalogue number (of a Schubert work).}
    \label{fig:midi}
\end{figure}

\subsubsection{Neuronal Partitioning}
Neural partitioning is run for the Melbourne daily minimum temperature task with results reported in Figure \ref{fig:melbourne_budget}. The trends of $N_R$ and $N_L$ are less apparent than that observed with the Mackey Glass forecasting task. The gradient of error is consistent with the change in $N_R$ and $N_L$ for deeper topologies between the tasks, but the same does not hold for broader networks; in fact, the inverse is observed, which suggests that hierarchical features and larger memory capacity is required to improve performance. Elongated memory capacity has been shown to emerge with the depth of an ESN \cite{gallicchioLocalLyapunovExponents2018}, which supports this observation.

\subsection{Polyphonic Music Tasks}
We evaluate the \emph{Mod-DeepESN} on a set of polyphonic music tasks as defined in \cite{boulanger2012modeling}. In particular, we use the data
provided\footnote{\url{http://www-etud.iro.umontreal.ca/~boulanni/icml2012}}
for the
Piano-midi.de\footnote{Classical MIDI piano music (\url{http://piano-midi.de/}).}
task. The data comprises a set of piano roll sequences preprocessed as described in \cite{poliner2006}. 87 sequences with an average of 872.5 timesteps are used for training, 12 sequences with an average of 711.7 timesteps are used for validation, and 25 sequences with an average of 761.4 timesteps are used for testing. The goal of this task is to predict $\mathbf{y}(t) = \mathbf{u}(t+1)$ given $\mathbf{u}(t)$ where $N_Y = N_U = 88$.

Multiple notes may be played at once, so an argmax cannot be used at the output of the readout layer; rather, the output of each neuron is binarized with a threshold. In practice, we find this threshold using training data by sweeping $\sim$20 values uniformly distributed between the minimum and maximum of the predicted values. This threshold can also be found by training a linear classifier on the predicted outputs, using an adaptive threshold, or using Otsu's method \cite{otsu1979threshold}. Lastly, an optimal threshold may exist on a per-neuron basis. A washout period of 20 steps is used to rid transients.

\begin{table}[t]
  \centering
  \caption{Piano-midi.de Time Series 1-Step Ahead Prediction Results.}
  \label{tab:piano_results}
  \ra{1.1}
  \begin{tabular}{@{}*{2}{c}@{}}
    \toprule
    Network & FL-ACC\\
    \midrule
    DeepESN  \cite{gallicchio_deep_2018}
    & 33.22\% \\
    shallowESN \cite{gallicchio_deep_2018}
    & 31.76\% \\
    RNN-RBM	     \cite{boulanger2012modeling}
    & 28.92\% \\
    \textit{Mod-DeepESN} & \textbf{33.44}\% \\
    \bottomrule
  \end{tabular}
\end{table}

\begin{figure*}
    \centering
    \begin{subfigure}{.05\linewidth}
      \centering
      \caption{}
    \end{subfigure}%
    \begin{subfigure}{.45\linewidth}
      \centering
      \includegraphics[width=.85\linewidth]{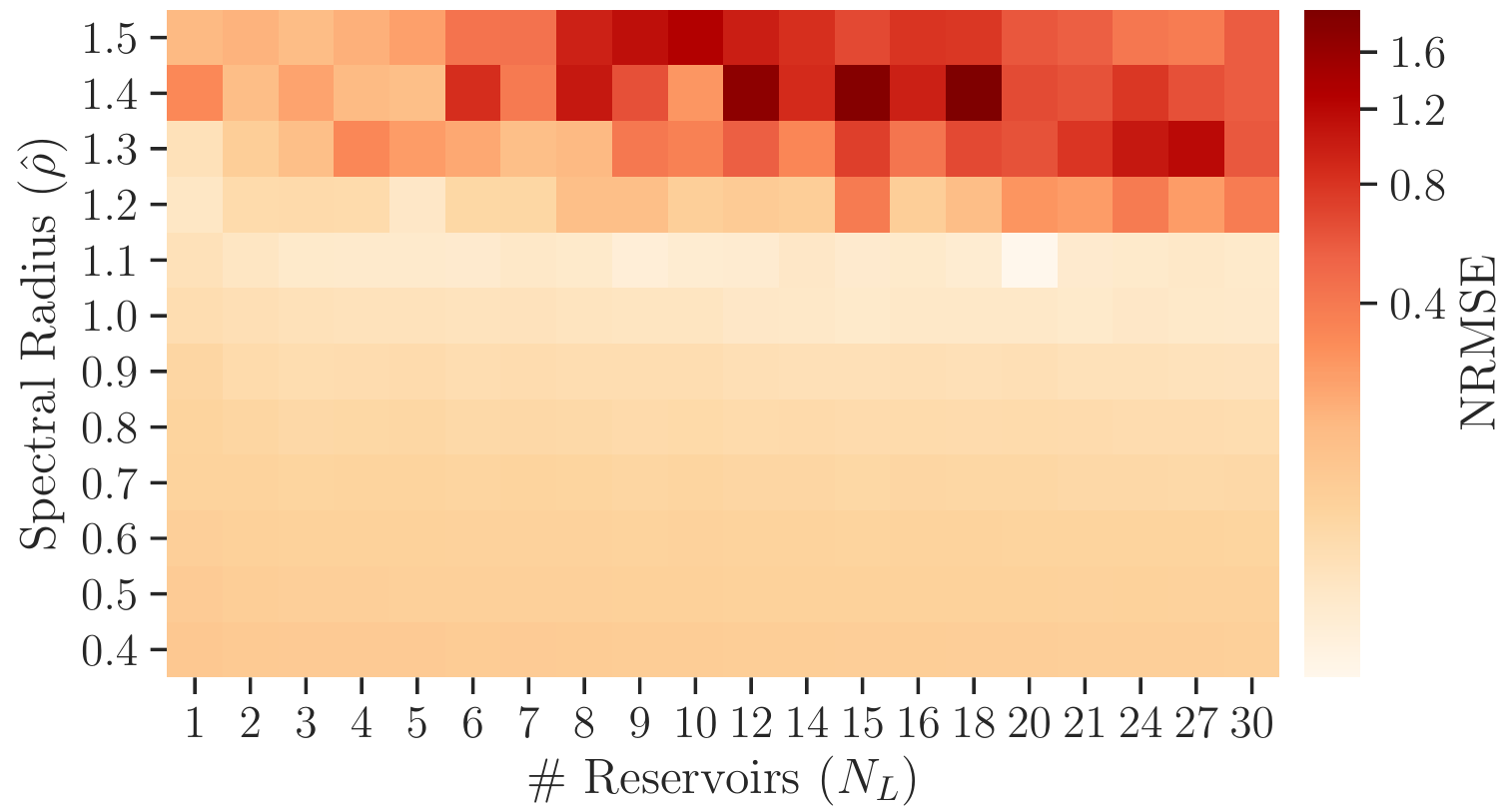}
      \label{fig:sweep_radius_hm}
    \end{subfigure}%
    \begin{subfigure}{.05\linewidth}
      \centering
      \caption{}
    \end{subfigure}%
    \begin{subfigure}{.45\linewidth}
      \centering
      \includegraphics[width=.85\linewidth]{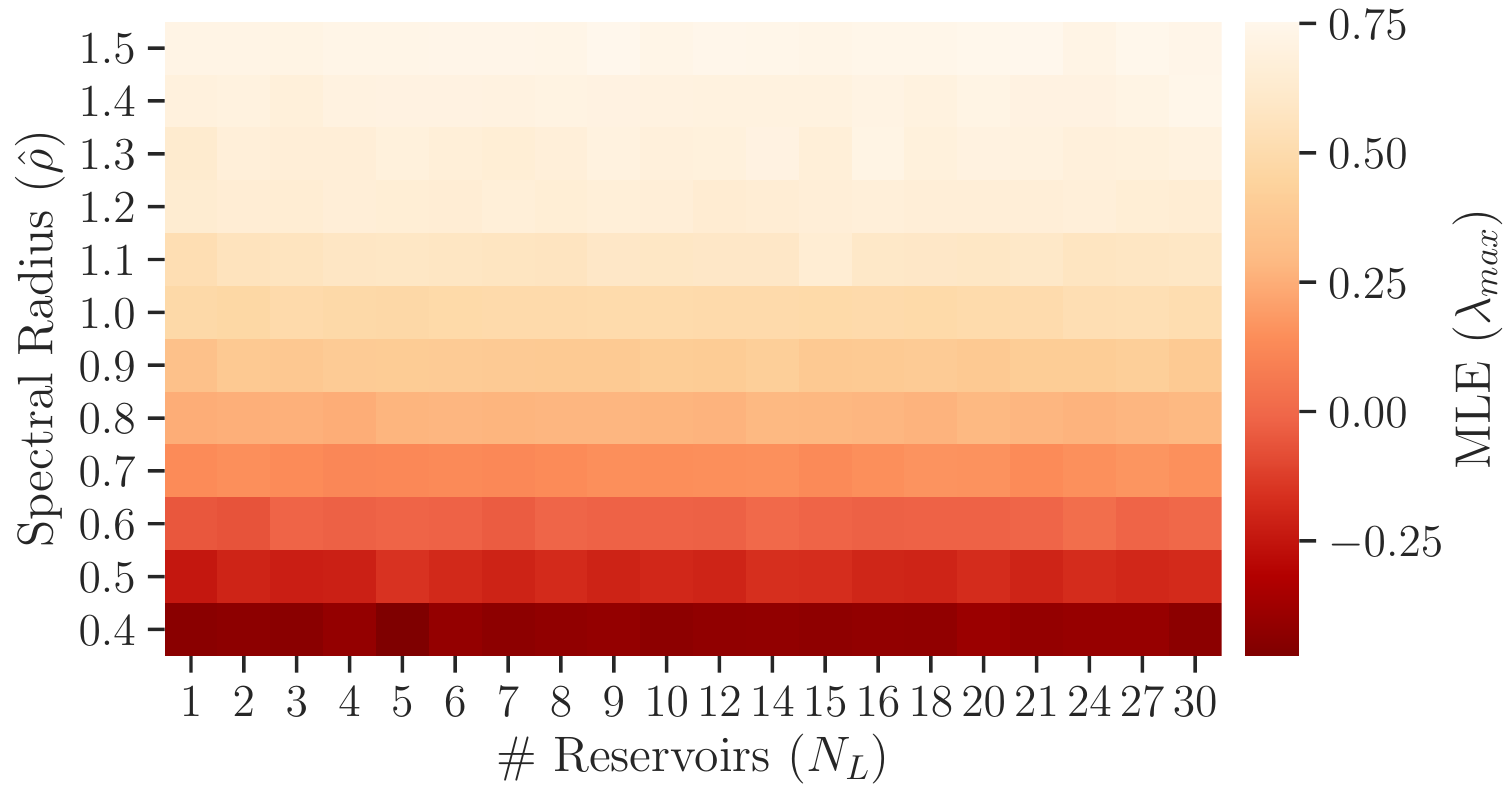}
      \label{fig:sweep_radius_hm_lambda}
    \end{subfigure}\vskip\baselineskip\vspace{-3mm}
    \begin{subfigure}{.05\linewidth}
      \centering
      \caption{}
    \end{subfigure}%
    \begin{subfigure}{.45\linewidth}
      \centering
      \includegraphics[width=.85\linewidth]{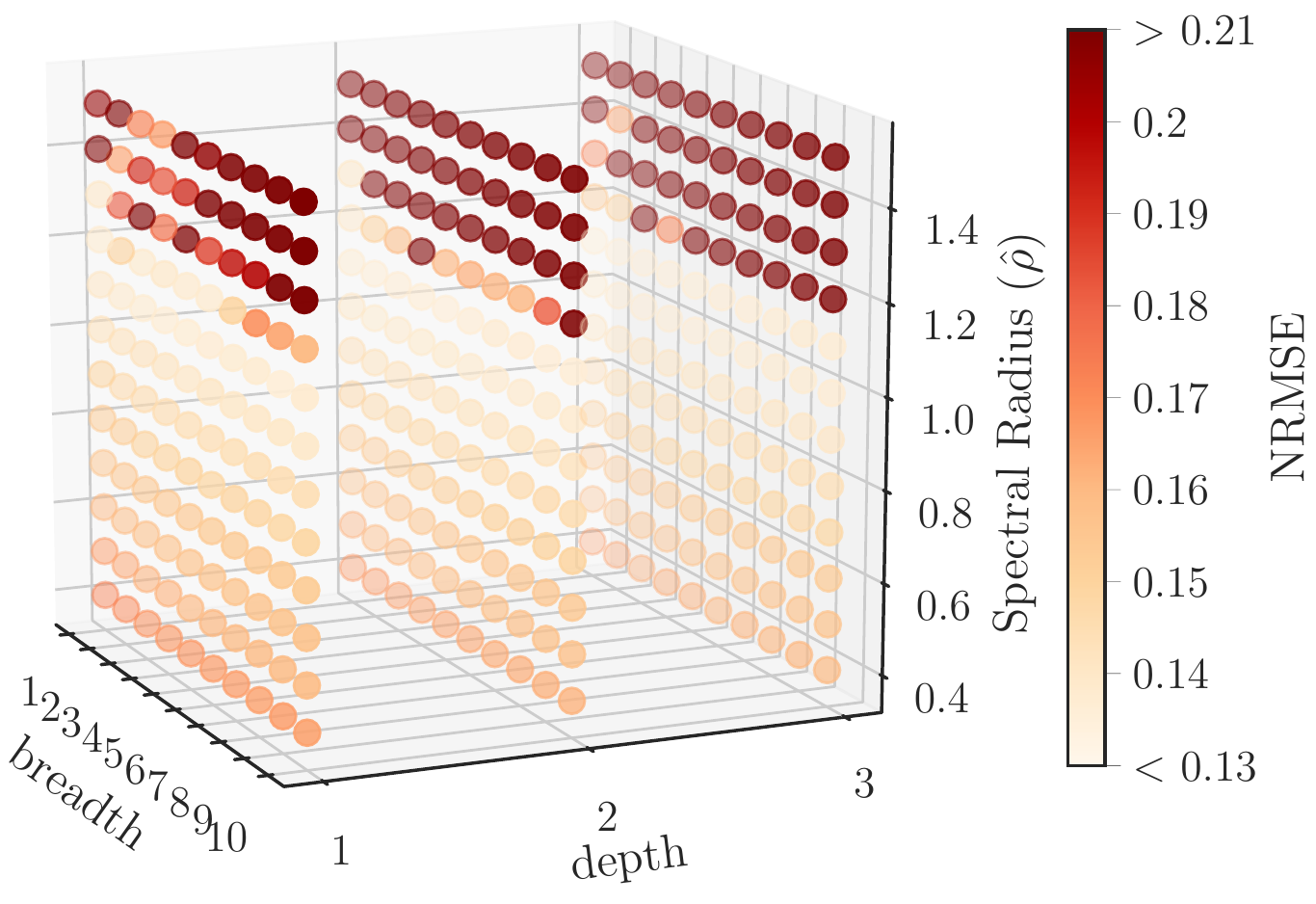}
      \label{fig:sweep_radius_3d}
    \end{subfigure}%
    \begin{subfigure}{.05\linewidth}
      \centering
      \caption{}
    \end{subfigure}%
    \begin{subfigure}{.45\linewidth}
      \centering
      \includegraphics[width=.85\linewidth]{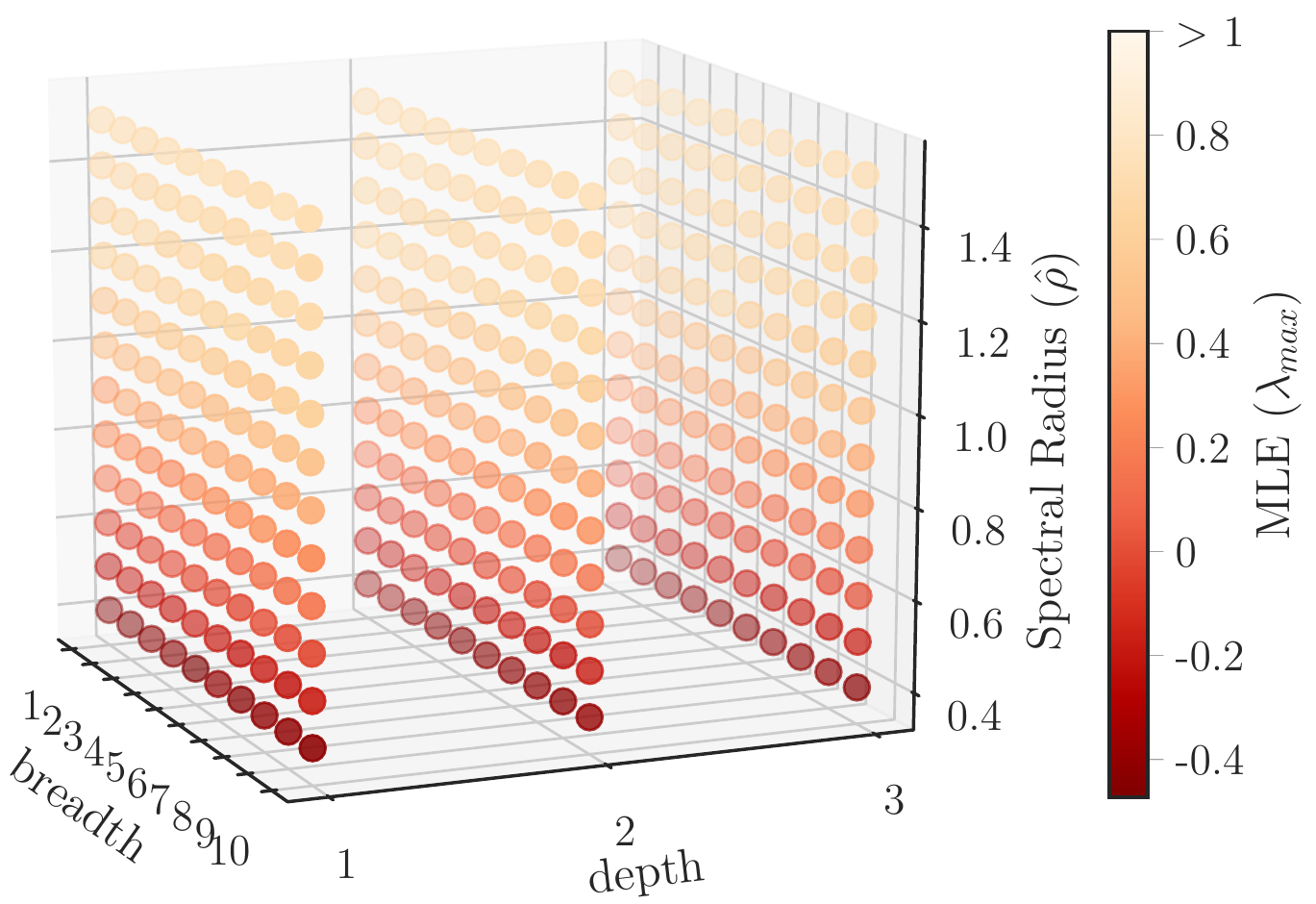}
      \label{fig:sweep_radius_3d_lambda}
    \end{subfigure}%
    \caption{2D and 3D heatmaps of NRMSE and $\lambda_{max}$ as a function of $\hat{\rho}$ and $N_L$. Note that the color bar gradient is reversed for visualizing $\lambda_{max}$. (a) Impact of $N_L$ and $\hat{\rho}$ on NRMSE\protect\footnotemark. (b) Impact of $N_L$ and $\hat{\rho}$ on $\lambda_{max}$. (c) Impact of reservoir breadth, depth, and $\hat{\rho}$ on NRMSE. NRMSE values span far greater than 0.21 and thus are clipped. (d) Impact of reservoir breadth, depth, and $\hat{\rho}$ on $\lambda_{max}$.}
    \label{fig:sweep_radius}
\end{figure*}

Table \ref{tab:piano_results} contains the best forecasting results for \emph{Mod-DeepESN} as well as those reported in \cite{boulanger2012modeling,gallicchio_deep_2018}. The \emph{Mod-DeepESN} framework outperforms both RNN-RBM \cite{boulanger2012modeling} and DeepESN \cite{gallicchio_deep_2018} on the Piano-midi.de corpus with fewer trained parameters and reservoirs. 


\section{Discussion}

It is evident that the optimal \emph{Mod-DeepESN} topologies found via neural partitioning are task-specific; no one configuration seems optimal for multiple tasks. Between the Mackey Glass and Melbourne forecasting tasks, wider networks exhibit a smaller confidence interval which indicates consistency in performance. There is less of an apparent trend on the real-world Melbourne task (more so for deep networks), although this is within expectations due to noise in the data.

\footnotetext{The colors of the color bar are linearly mapped to the interval $[0, 1]$ with power-law normalization, i.e. {\normalfont
    $\frac{
        \left( \mathbf{c} - \mathbf{c}_{\textup{min}} \right) ^ {\gamma}
    }{
        \left( \mathbf{c}_{\textup{max}} - \mathbf{c}_{\textup{min}} \right) ^ {\gamma}
    }$} ($\gamma = 0.3$) for some given colors, $\mathbf{c}$.}

We create separation ratio plots of reservoir responses to time series input with various examples shown in Figure \ref{fig:sep_ratios}. The best-performing models achieve similar error values, but exhibit considerably different dynamics at different scales of magnitude. The worst-performing model yields a separation ratio similar to that of the second-best; the biases differ, however, this can be attributed to the difference in magnitude (as a result of e.g. input scaling).
Looking at \eqref{eq:sep_ratio}, it can be observed that the identity function yields an ideal response at the empirical ``edge of chaos;'' this sheds light on some shortcomings of the metric. The technique can be made more robust by considering input-to-output similarity, matching the variance of inputs and reservoir responses (to avoid skewing the slope), and tracking the consistency of separation ratios over time (as reservoirs are stateful).
We recommend these plots as a debugging method for ESNs as they unveil useful attributes beyond input and output separation.

Of the three tasks considered, the Melbourne daily minimum forecasting task is selected for exploring the design space. The data is non-synthetic and its size is not prohibitive of such exploration. Here, we produce heatmaps of NRMSE and MLE ($\lambda_{max}$) as a function of several swept hyperparameters. In each, a practical range of breadth and depth values are considered.
Figure \ref{fig:sweep_alpha} delineates the impact of $\alpha$ and shows that $\lambda_{max}$ is a reliable indication of performance ($\rho = -0.956$%
\footnote{Pearson correlation coefficient \cite{pearson1895} between NRMSE and $\lambda_{max}$ (do not confuse with $\hat{\rho}$, the spectral radius).})
for the task. There is no significant impact induced by modulating breadth or depth, which agrees with the neuronal partitioning result (see Figure \ref{fig:melbourne_budget}).
Figure \ref{fig:sweep_radius} illustrates the impact of $\hat{\rho}$ and demonstrably has a more substantial impact on network stability than $\alpha$, as expected. The network error plateaus to a minimum near \SI{1.1} and increases dramatically afterward. This critical point is beyond the ``edge of chaos'' ($\lambda_{max} = 0$) and error is asymmetrical about it. Here, $\lambda_{max}$ is a poor predictor of error ($\rho = 0.377$) (even the correlation between $\hat{\rho}$ and NRMSE is higher). Again, depth and breadth are not indicative of error on their own, however, both deeper and wider networks suffer larger errors beyond the critical value of $\hat{\rho}$.

An interesting observation is that, while the tasks differ, well-performing networks in this work often exhibit a positive $\lambda_{max}$ whereas the networks in \cite{gallicchioLocalLyapunovExponents2018} exhibit a negative $\lambda_{max}$ primarily. This characterization of \emph{Mod-DeepESN} as a system with ``unstable'' dynamics requires further attestation but indicates that such does not preclude consistent performance.


\section{Conclusion}

We provide analytical rationale for the characteristics of deep ESN design that influence forecasting performance. Within the malleable \emph{Mod-DeepESN} architecture, we experimentally support that networks perform optimally beyond the ``edge of chaos.'' Provided constraints on model size or compute resources, we explore the effects of neuron allocation and reservoir placement on performance. We also demonstrate that network breadth plays a role in dictating certainty of performance between instances. These characteristics may be present within neuronal populations, which could confirm that powerful models emerge from an agglomeration of weak learners. Redundancy through parallel pathways, extraction of nonlinear data regularities with depth, and discernibility of latent representations all appear to have a significant impact on \emph{Mod-DeepESN} performance. Future studies should explore the design space of reservoirs in tandem with neuromorphic hardware design constraints.

\begin{acks}
    We would like to thank the members of the Neuromorphic Artificial Intelligence Lab for helpful discussions during this work. We also thank the creators and maintainers of the \verb|matplotlib| \cite{matplotlib} and \verb|seaborn| \cite{seaborn} libraries which we used for plotting.
    
    We acknowledge the Air Force Research Lab in funding part of this work under agreement number FA8750-16-1-0108. The U.S. Government is authorized to reproduce and distribute reprints for Governmental purposes notwithstanding any copyright notation thereon.
    The views and conclusions contained herein are those of the authors and should not be interpreted as necessarily representing the official policies or endorsements, either expressed or implied, of Air Force Research Laboratory or the U.S. Government.
\end{acks}

\begin{figure*}[!b]
    \centering
    \begin{subfigure}{.05\linewidth}
      \centering
      \caption{}
      \label{fig:sweep_sparsity_l_hm}
    \end{subfigure}%
    \begin{subfigure}{.45\linewidth}
      \centering
      \includegraphics[width=.85\linewidth]{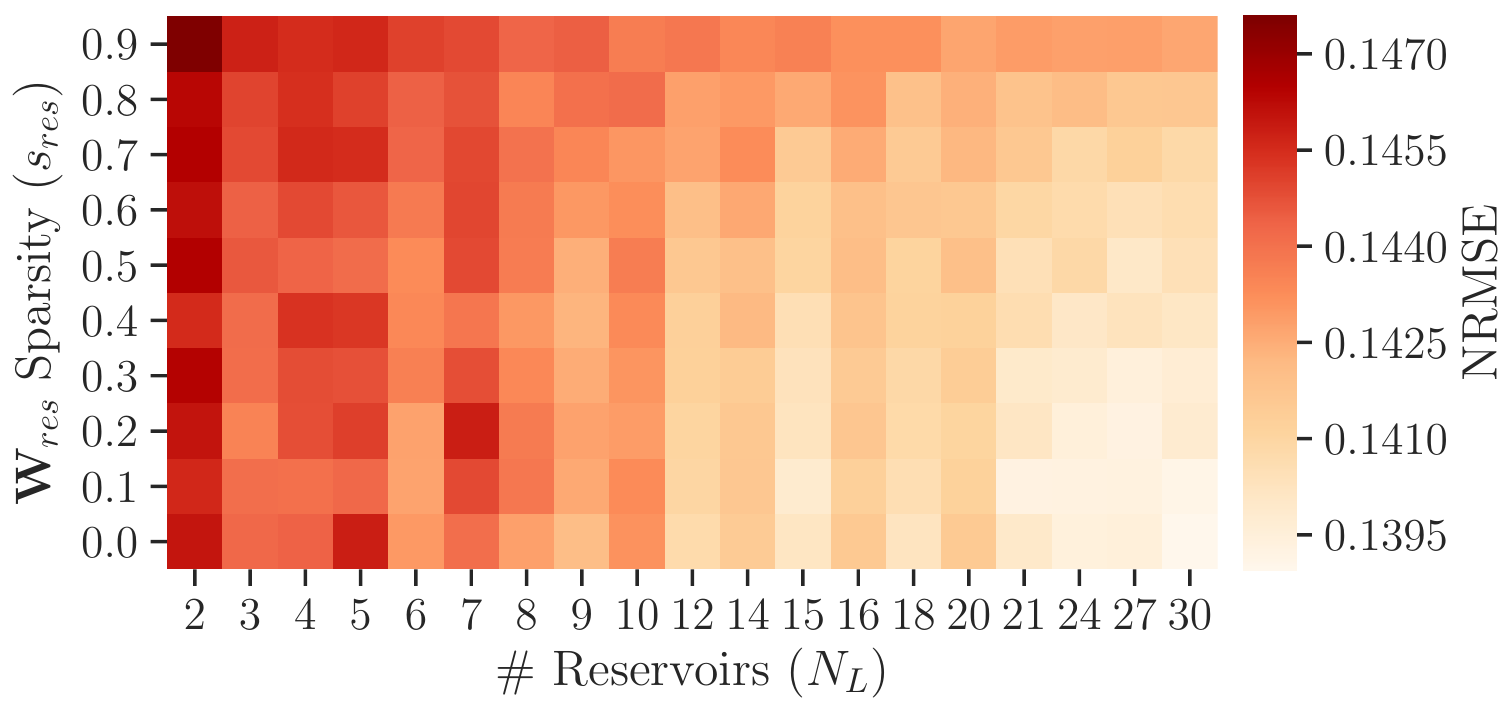}
    \end{subfigure}%
    \begin{subfigure}{.05\linewidth}
      \centering
      \caption{}
      \label{fig:sweep_sparsity_l_hm_lambda}
    \end{subfigure}%
    \begin{subfigure}{.45\linewidth}
      \centering
      \includegraphics[width=.85\linewidth]{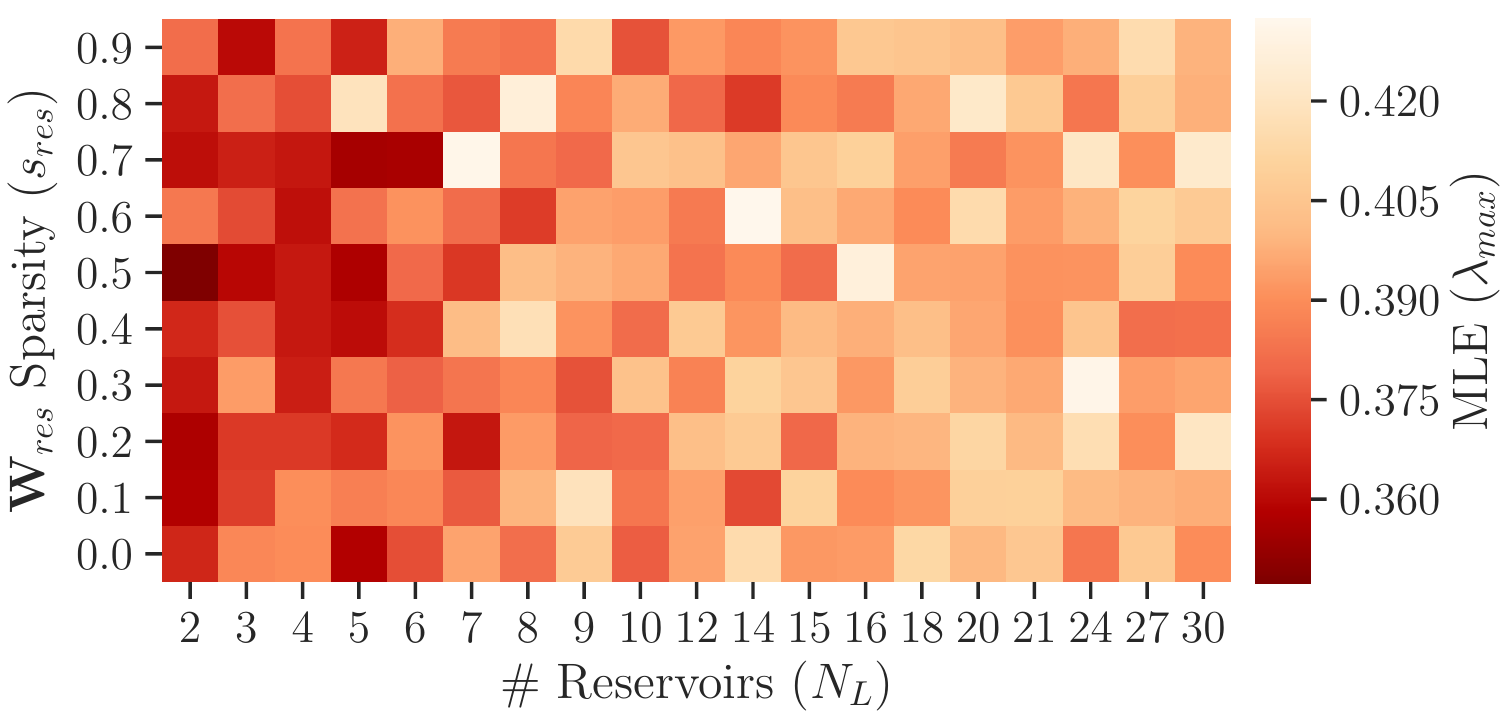}
    \end{subfigure}\vskip\baselineskip\vspace{-3mm}
    \begin{subfigure}{.05\linewidth}
      \centering
      \caption{}
      \label{fig:sweep_sparsity_l_3d}
    \end{subfigure}%
    \begin{subfigure}{.45\linewidth}
      \centering
      \includegraphics[width=.85\linewidth]{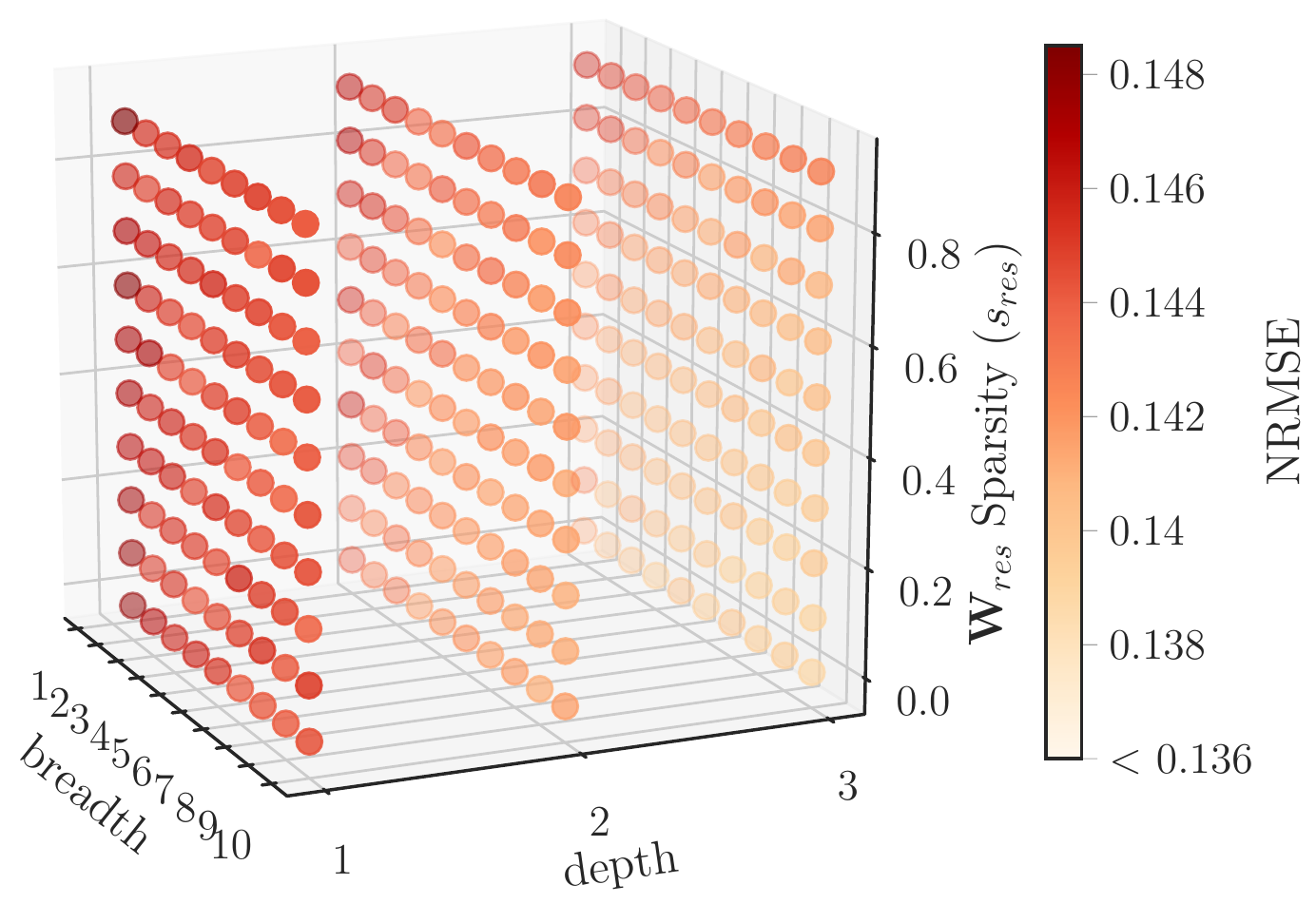}
    \end{subfigure}%
    \begin{subfigure}{.05\linewidth}
      \centering
      \caption{}
      \label{fig:sweep_sparsity_l_3d_lambda}
    \end{subfigure}%
    \begin{subfigure}{.45\linewidth}
      \centering
      \includegraphics[width=.85\linewidth]{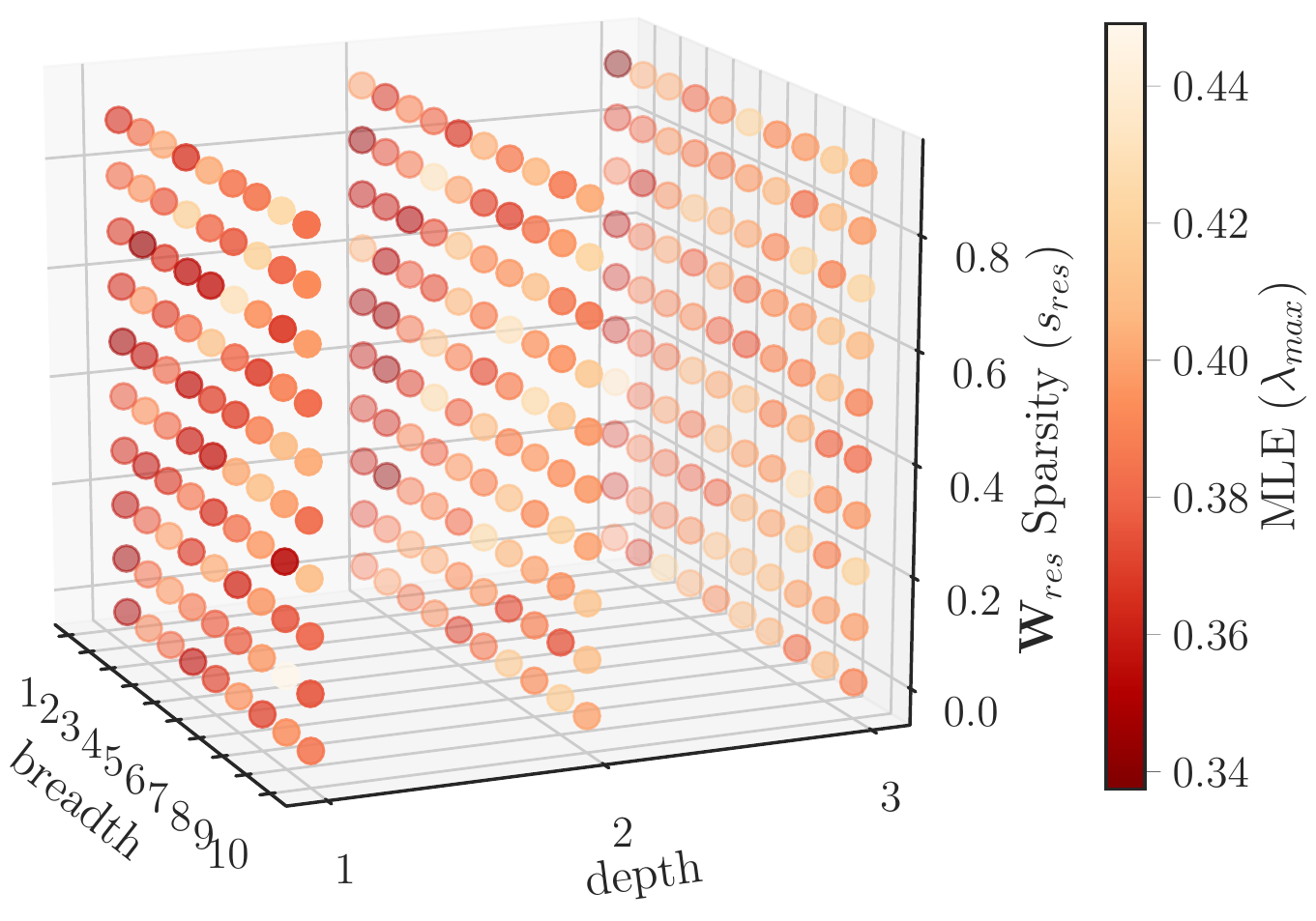}
    \end{subfigure}%
    \caption{2D and 3D heatmaps of NRMSE and $\lambda_{max}$ as a function of $s_{res}$ and $N_L$. Note that the color bar gradient is reversed for visualizing $\lambda_{max}$. (a) Impact of $N_L$ and $s_{res}$ on NRMSE. (b) Impact of $N_L$ and $s_{res}$ on $\lambda_{max}$. (c) Impact of reservoir breadth, depth, and $s_{res}$ on NRMSE. (d) Impact of reservoir breadth, depth, and $s_{res}$ on $\lambda_{max}$.}
    \label{fig:sweep_sparsity_l}
\end{figure*}

\bibliographystyle{ACM-Reference-Format}
\bibliography{acmart}

\begin{appendix}

\section{Additional Results}
We additionally construct heatmaps for the impact of topology and $s_{res}$ on NRMSE and $\lambda_{max}$, shown in Figure \ref{fig:sweep_sparsity_l}. This result shows that $\lambda_{max}$ somewhat correlates with NRMSE ($\rho = -0.544$), which only moderately supports the ``edge of chaos'' hypothesis. However, there is a clear trend between NRMSE and both $s_{res}$ and depth. The error extending outward radially from the bottom-right corner of Figure \ref{fig:sweep_sparsity_l_hm} correlates positively with decreasing $s_{res}$ and increasing $N_L$. More significantly, depth correlates with NRMSE ($\rho = -0.749$) with deeper networks giving lower errors. Wider networks also always yield lower errors in this experiment.

\end{appendix}

\end{document}